**Spectrophotometry of (32) Pomona, (145) Adeona, (704) Interamnia, (779) Nina, (330825) 2008 XE3, and 2012 QG42 and laboratory study of possible analog samples**

Vladimir V. Busarev[a, b, *], Sergey I. Barabanov[b], Vyacheslav S. Rusakov[c], Vasiliy B. Puzin[b], Valery V. Kravtsov[a, d]

[a]*Lomonosov Moscow State University, Sternberg Astronomical Institute, University Avenue 13, 119992 Moscow, Russia*

[b]*Institute of Astronomy of Russian Academy of Science, Pyatnitskaya St. 48, 109017 Moscow, Russia*

[c]*Division of Mossbauer Spectroscopy, Physical Dep. of Lomonosov Moscow State University, 119992 Moscow, Russia*

[d]*Departamento de Fisica, Facultad de Ciencias Naturales, Universidad de Atacama, Copayapu 485, Copiapo, Chile*

[*] Corresponding author.

E-mail address: busarev@sai.msu.ru (V.V. Busarev)

**Abstract**

Six asteroids including two NEAs, one of which is PHA, accessible for observation in September 2012 were investigated using a low-resolution (R ≈ 100) spectrophotometry in the range 0.35-0.90 μm with the aim to study features of their reflectance spectra. A high-altitude position of our Terskol observatory (3150 m above sea level) favorable for the near-UV and visible-range observations of celestial objects let us probably to detect some new spectral features of the asteroids. Two subtle absorption bands centered at 0.53 and 0.74 μm were found in the reflectance spectra of S-type (32) Pomona and interpreted as signs of presence of pyroxenes in the asteroid surface matter and its different oxidation. Very similar absorption bands centered at 0.38, 0.44 and 0.67-0.71 μm have been registered in the reflectance spectra of (145) Adeona, (704) Interamnia, and (779) Nina of primitive types. We performed laboratory investigations of ground samples of known carbonaceous chondrites, Orguel (CI), Mighei (CM2), Murchison (CM2), Boriskino (CM2), and seven samples of low-iron Mg serpentines as possible analogs of the primitive asteroids. In the course of this work, we discovered an intense absorption band (up to ~25%) centered at 0.44 μm in reflectance spectra of the low-Fe serpentine samples. As it turned out, the equivalent width of the band has a high correlation with content of $Fe^{3+}$ (octahedral and tetrahedral) in the samples. It may be considered as a confirmation of the previously proposed mechanism of the absorption due to electronic transitions in exchange-coupled pairs (ECP) of $Fe^{3+}$neighboring cations. It means that the absorption feature can be used as an indicator of ferric iron in oxidized and hydrated low-Fe compounds on the surface of asteroids and other atmosphereless celestial bodies. Moreover, our measurements showed that the mechanism of light absorption is partially or completely blocked in the case of intermediate to high iron contents. Therefore, the method cannot probably be used for quantitative estimation of $Fe^{3+}$ content on the bodies. Based on laboratory study of the analog samples, we conclude that spectral characteristics of Adeona, Interamnia, and Nina correspond to a mixture of CI-CM-chondrites and hydrated silicates, oxides and/or hydroxides. Spectral signs of sublimation activity on Adeona, Interamnia, and Nina at minimal heliocentric distances are likely discovered in the short-wavelength range (~0.4-0.6 μm). It is suggested that such cometary-like activity at the highest surface temperatures is a frequent phenomenon for C and close type asteroids including considerable amounts of ices beneath the surface. A usual way of origin of a temporal coma of ice particles around a primitive asteroid is excavated fresh ice at recent impuct event(s).

The obtained reflectance spectra of two NEAs, (330825) 2008 XE3 and 2012 QG42, are predominantly featureless and could be attributed to S(C) and S(B)-type bodies, respectively. We discuss reasons why weak spectral features seen in reflectance spectra of the main-belt asteroids are not observed in those of NEAs.

## 1. Introduction

Spectrophotometry/ spectroscopy is a traditional method of remote study of asteroids and other atmosphereless celestial bodies (e. g., McCord et al., 1970; Adams, 1974). When ground-based telescopes used, the range preliminary from ~0.38 to 1.1 microns is extended up to 2.5 microns for the last two decades (e. g., Vernazza et al., 2008; DeMeo et al., 2009, Hardersen et al., 2014; Fieber-Beyer et al., 2015). It is defined by the boundaries of the most transparent spectral "window" of the Earth's atmosphere, through which the bulk of observational information on asteroids was obtained. It allowed us to enrich our knowledge about these objects, in particular on their taxonomy (e.g., Tholen, 1989; Bus and Binzel, 2002 a,b; DeMeo et al., 2009). Further progress in ground-based studies of asteroids are naturally related with increasing both the number (and therefore the sample size) of studied bodies and the accuracy of spectral measurements. High-quality reflectance spectra of asteroids potentially contain not only valuable mineralogical information on the material of which the asteroids are made but also that on the valency state of iron (as well as of other transition metals). As is known, the latter depends on physico-chemical parameters (Platonov, 1976; Burns, 1993) of asteroid matter connected with the formation conditions of the bodies and their subsequent evolution. Unfortunately, various distorting factors, such as observational faults and space weathering, make it difficult to reconstruct reliably the previous conditions. Thus, the final goal of such kind of study is to accurately extract the observational information and to try to correctly interpret it. For most asteroids, except for some bodies investigated by space methods, there is a lack of data about whether or not chemico-mineralogical and other properties vary along their surface. The reason of that is mainly due to (nearly) point-like appearance of asteroids at ground-based observations, which makes it difficult to obtain spectral information of different parts of the asteroids' surface. Indeed, the angular size of (1) Ceres, the largest asteroid with a diameter of ~ 1000 km, varies in the range 0.″8−0.″3, which is comparable to the limiting angular resolution of ground-based telescopes at the excellent atmospheric seeing. The most important is to minimize the impact of the Earth's atmosphere on reflectance spectra of asteroids and on the reliability of final results and conclusions. To achieve this goal it is useful to compare spectral data obtained with the same facility on asteroids of the same and/or close taxonomic types which are expected to have similar spectral features. Such approach is used in the work to study several of C-B-type asteroids. As before, a laboratory study of spectral characteristics of both meteoritic and terrestrial analog samples is very helpful and used also in our work. In addition, we consider and discuss the reflectance spectra obtained by other authors for the objects of our sample to the date.

## 2. Observations and data reduction

Spectrophotometry of the asteroids (32) Pomona, (145) Adeona, (704) Interamnia, (779) Nina, (330825) 2008 XE3, and 2012 QG42 was performed in September 2012 using the 2-m telescope of Terskol Observatory operated by Institute of Astronomy of Russian Academy of Science (IA RAS). The observatory is situated at high altitude of 3150-m above sea level, making especially favorable conditions for observations at shorter wavelengths. The telescope is equipped with a prism CCD-spectrometer (WI CCD 1240x1150 pix.) working in the range 0.35-0.97 μm, with $R \approx 100$ resolving power. DECH spectral package (Galazutdinov, 1992) was employed to reduce CCD observations by means of standard reduction procedures (such as flat-field correction, bias and dark subtraction, etc.) and to extract asteroid spectra. Wavelength calibration of the spectra was done using the positions of hydrogen Balmer lines in the spectrum of α Peg (B9III) observed in a repeated mode. The total exposure time spent on each target was typically ~1-2$^h$. The obtained reflectance spectra were corrected for the difference in air mass by applying a conventional method based on using observations of a solar analog star (e.g., McCord et al., 1970). In our work a single solar analog star, HD 10307 (G1.5V) (Hardorp, 1980), was intentionally exploited to avoid possible differences in the calculated reflectance spectra of asteroids as in the case of several solar analogs use. Observations of the same star were performed to determine the running spectral extinction function of the terrestrial atmosphere (Busarev, 2011). The observations of HD 10307 were made nearly in the same range of the air masses at (or close to) which the asteroids of the sample were observed (see Table 1). The values of the signal-to-noise ratio (S/N) of the asteroid spectra were estimated in the range of 0.4-0.8 μm. They are given, along with other data, in Table 1. To reduce high-frequency fluctuations in the reflectance spectra, they were smoothed by the method of ”running box average” with a 5-point averaging interval. This allowed us to study considerably wider spectral features of the observed asteroids and to assess their spectral types according to shapes of their reflectance spectra. As a rule, averaging of the asteroid consecutive reflectance spectra was made when they had a close overall shape and observational spectra were obtained at minimal air masses.

Ephemerides (taken from the IAU Minor Planet Center on-line service at http://www.minorplanetcenter.net/iau/MPEph/MPEph.html) and observation parameters of the asteroids are given in Table 1.

Table 1. Ephemerides and observational parameters of asteroids for moments of all obtained spectra.

| 32 Pomona | | | | | | | | | | | | |
|---|---|---|---|---|---|---|---|---|---|---|---|---|
| Date (y m d) | UT meadle (hms) | R.A. (h m s) | Decl. (° ′ ) | Delta (AU) | r (AU) | Elong.(°) | Ph.(°) | V($^m$) | Elev.(°) | Exp. Time (s) | Air Mass | S/N |
| 2012 09 19 | 234100 | 02 01.01 | +15 10.0 | 1.926 | 2.799 | 143.7 | 12.3 | 12.0 | 61.2 | 1200 | 1.1434 | 9.3 |
| 2012 09 20 | 000200 | 02 01.01 | +15 09.9 | 1.925 | 2.799 | 143.8 | 12.2 | 12.0 | 60 | 1200 | 1.1543 | 7.2 |
| 2012 09 20 | 002300 | 02 01.00 | +15 09.9 | 1.925 | 2.799 | 143.8 | 12.2 | 12.0 | 58 | 1200 | 1.1788 | 6.2 |
| 2012 09 20 | 004300 | 02 00.99 | +15 09.8 | 1.925 | 2.799 | 143.8 | 12.2 | 12.0 | 57 | 1200 | 1.2062 | 10 |
| 2012 09 20 | 005400 | 02 00.99 | +15 09.8 | 1.925 | 2.799 | 143.8 | 12.2 | 12.0 | 55 | 600 | 1.2202 | 7.6 |
| 2012 09 20 | 010500 | 02 00.98 | +15 09.8 | 1.925 | 2.799 | 143.8 | 12.2 | 12.0 | 53 | 600 | 1.2515 | 7.5 |
| 2012 09 20 | 011600 | 02 00.98 | +15 09.7 | 1.925 | 2.799 | 143.8 | 12.2 | 12.0 | 51.5 | 600 | 1.2770 | 6.3 |
| 2012 09 20 | 012700 | 02 00.98 | +15 09.7 | 1.925 | 2.799 | 143.8 | 12.2 | 12.0 | 50 | 600 | 1.3046 | 6 |
| Solar analog star HD10307: Elev. 61$^{\circ}$ t=01$^h$ 30$^m$, Air Mass 1.1430 | | | | | | | | | | | | |
| | | | | | | | | | | | | |
| 145 Adeona | | | | | | | | | | | | |
| Date (y m d) | UT meadle ( hms) | R.A. (h m s) | Decl. (° ′ ) | Delta (AU) | r (AU) | Elong.(°) | Ph.(°) | V($^m$) | Elev.(°) | Exp. Time (s) | Air Mass | S/N |
| 2012 09 19 | 225400 | 02 29.58 | -01 38.1 | 1.843 | 2.692 | 140.1 | 13.9 | 12.4 | 43.5 | 1200 | 1.4512 | 3 |
| 2012 09 19 | 231500 | 02 29.57 | -01 38.1 | 1.843 | 2.692 | 140.1 | 13.9 | 12.4 | 44.5 | 1200 | 1.4253 | 3.2 |
| Solar analog star HD10307: Elev. 67$^{\circ}$ t=01$^h$ 32$^m$ , Air Mass 1.0862 | | | | | | | | | | | | |
| | | | | | | | | | | | | |
| 704 Interamnia | | | | | | | | | | | | |
| Date (y m d) | UT meadle ( hms ) | R.A. (h m s) | Decl. (° ′ ) | Delta (AU) | r (AU) | Elong.(°) | Ph.(°) | V($^m$) | Elev.(°) | Exp. Time (s) | Air Mass | S/N |
| 2012 09 13 | 211100 | 03 24.64 | +39 00.2 | 2.080 | 2.616 | 111.1 | 21.0 | 10.9 | 47 | 600 | 1.3662 | 5 |
| 2012 09 13 | 212200 | 03 24.64 | +39 00.2 | 2.080 | 2.616 | 111.1 | 21.0 | 10.9 | 49 | 600 | 1.3242 | 5.1 |
| 2012 09 13 | 213300 | 03 24.64 | +39 00.3 | 2.080 | 2.616 | 111.1 | 21.0 | 10.9 | 51 | 600 | 1.2860 | 4.9 |
| 2012 09 13 | 214400 | 03 24.65 | +39 00.3 | 2.080 | 2.616 | 111.1 | 21.0 | 10.9 | 53 | 600 | 1.2515 | 5 |
| 2012 09 13 | 215500 | 03 24.65 | +39 00.4 | 2.080 | 2.616 | 111.1 | 21.0 | 10.9 | 55 | 600 | 1.2202 | 5 |
| Solar analog star HD10307: Elev. 48$^{\circ}$ at t=19$^h$ 25$^m$ , Air Mass 1.3446 | | | | | | | | | | | | |
| | | | | | | | | | | | | |
| 779 Nina | | | | | | | | | | | | |
| Date (y m d) | UT meadle ( hms ) | R.A. (h m s) | Decl. (° ′ ) | Delta (AU) | r (AU) | Elong.(°) | Ph.(°) | V($^m$) | Elev.(°) | Exp. Time (s) | Air Mass | S/N |
| 2012 09 13 | 175100 | 00 53.11 | +32 40.7 | 1.289 | 2.149 | 138.5 | 18.1 | 11.0 | 35.5 | 300 | 1.7190 | 4.4 |
| 2012 09 13 | 175700 | 00 53.11 | +32 40.7 | 1.289 | 2.149 | 138.5 | 18.1 | 11.0 | 36.5 | 300 | 1.6783 | 4.5 |
| 2012 09 13 | 180300 | 00 53.10 | +32 40.7 | 1.289 | 2.149 | 138.5 | 18.1 | 11.0 | 37.5 | 300 | 1.6401 | 4.6 |
| 2012 09 13 | 180900 | 00 53.10 | +32 40.7 | 1.289 | 2.149 | 138.5 | 18.1 | 11.0 | 38.5 | 300 | 1.6040 | 4.6 |
| 2012 09 13 | 181500 | 00 53.10 | +32 40.7 | 1.289 | 2.149 | 138.5 | 18.1 | 11.0 | 39.5 | 300 | 1.5700 | 5.5 |
| 2012 09 13 | 201300 | 00 53.04 | +32 41.0 | 1.289 | 2.149 | 138.6 | 18.0 | 11.0 | 61.5 | 300 | 1.1376 | 5.5 |
| 2012 09 13 | 201900 | 00 53.04 | +32 41.0 | 1.289 | 2.149 | 138.6 | 18.0 | 11.0 | 62.5 | 300 | 1.1271 | 5.3 |
| 2012 09 13 | 202500 | 00 53.03 | +32 41.0 | 1.289 | 2.149 | 138.6 | 18.0 | 11.0 | 63.5 | 300 | 1.1171 | 5.4 |
| 2012 09 13 | 203100 | 00 53.03 | +32 41.1 | 1.288 | 2.149 | 138.6 | 18.0 | 11.0 | 64.5 | 300 | 1.1077 | 5.5 |
| 2012 09 13 | 203700 | 00 53.03 | +32 41.1 | 1.288 | 2.149 | 138.6 | 18.0 | 11.0 | 65.5 | 300 | 1.0987 | 4.3 |
| 2012 09 13 | 224400 | 00 52.97 | +32 41.4 | 1.288 | 2.149 | 138.7 | 18.0 | 11.0 | 79 | 900 | 1.0187 | 5.3 |
| 2012 09 13 | 230000 | 00 52.96 | +32 41.4 | 1.288 | 2.149 | 138.7 | 18.0 | 11.0 | 78 | 900 | 1.0223 | 5.1 |
| Solar analog star HD10307: Elev. 48$^{\circ}$ at t=19$^h$ 25$^m$, Air Mass 1.3446 | | | | | | | | | | | | |
| | | | | | | | | | | | | |
| (330825) 2008 XE3 | | | | | | | | | | | | |
| Date (y m d) | UT meadle ( hms ) | R.A. (h m s) | Decl. (° ′ ) | Delta (AU) | r (AU) | Elong.(°) | Ph.(°) | V($^m$) | Elev.(°) | Exp. Time (s) | Air Mass | S/N |
| 2012 09 12 | 223200 | 01 37 07.8 | +37 18.0 | 0.246 | 1.176 | 128.6 | 42.0 | 15.2 | 79.5 | 1800 | 1.0170 | 2 |
| 2012 09 12 | 230300 | 01 37 11.7 | +37 18.9 | 0.246 | 1.176 | 128.6 | 42.0 | 15.2 | 83 | 1800 | 1.0075 | 2 |
| 2012 09 12 | 233400 | 01 37 15.7 | +37 19.9 | 0.246 | 1.176 | 128.6 | 42.0 | 15.2 | 83 | 1800 | 1.0075 | 3.2 |
| 2012 09 13 | 003600 | 01 37 23.5 | +37 21.7 | 0.246 | 1.176 | 128.6 | 42.0 | 15.2 | 74 | 1800 | 1.0402 | 2.8 |
| 2012 09 13 | 010700 | 01 37 27.4 | +37 22.6 | 0.246 | 1.176 | 128.6 | 42.0 | 15.2 | 68.5 | 1800 | 1.0746 | 4.9 |
| Solar analog star HD10307: Elev. 67$^{\circ}$ at t=01$^h$ 32$^m$, Air Mass 1.0862 | | | | | | | | | | | | |
| | | | | | | | | | | | | |
| 2012 QG42 | | | | | | | | | | | | |
| Date (y m d) | UT meadle ( hms ) | R.A. (h m s) | Decl. (° ′ ) | Delta (AU) | r (AU) | Elong.(°) | Ph.(°) | V($^m$) | Elev.(°) | Exp. Time (s) | Air Mass | S/N |
| 2012 09 12 | 190700 | 20 35 40.4 | +10 22.8 | 0.021 | 1.021 | 135.7 | 43.5 | 14.2 | 55 | 1200 | 1.2202 | 9 |
| 2012 09 12 | 193200 | 20 34 38.3 | +10 28.7 | 0.021 | 1.021 | 135.4 | 43.7 | 14.3 | 53 | 1200 | 1.2515 | 7.4 |
| 2012 09 12 | 195300 | 20 33 46.0 | +10 33.6 | 0.021 | 1.021 | 135.2 | 44.0 | 14.3 | 51 | 1200 | 1.2860 | 6.4 |
| 2012 09 12 | 201400 | 20 32 53.4 | +10 38.6 | 0.021 | 1.021 | 135.0 | 44.2 | 14.3 | 48.5 | 1200 | 1.3342 | 6.8 |
| Solar analog star HD10307: Elev. 67$^{\circ}$ at t=01$^h$ 32$^m$, Air Mass 1.0862 | | | | | | | | | | | | |

## 3. Analysis and interpretation of asteroid reflectance spectra

### 3.1. 32 Pomona

Average diameter and geometric albedo of Pomona according to recent WISE-data are of 81.78 km and 0.25 (Masiero et al., 2014). The asteroid rotates with a period of 9.448$^h$ (Harris et al., 2012). In total, eight separate spectra of Pomona were registered on the night 19/20 of September, 2012, together with the spectra of HD10307 used as a solar analog star (Table 1). We used them to calculate an average spectrum of the asteroid and normalized it to 1.0 at 0.55 μm (Figure 1a). It corresponds to an S-type body having mineralogy dominated by pyroxenes, olivines, and other high-temperature compounds (e.g.,

Gaffey et al., 1989; Gaffey et al., 1993). Given the rotational period of the asteroid, the total exposure time spent to obtain the eight spectra (2 hours) corresponds to ≈ 1/5 of its rotational period. In terms of its shape, the spectrum is very similar to that obtained by McCord and Chapman (1975a) but differs to some extent from that by Bus and Binzel (2002a or 2003a) and Xu et al. (1995) (Figure 1b). We find two relatively weak and broad absorption bands in the obtained spectra of Pomona at 0.49-0.55 μm and 0.73-0.77 μm (Figure 1a). We suppose that the former one, at 0.49-0.55 μm, has a complex nature. It is probably a superposition of two or even three wide spectral features. It was established earlier that electronic spin-forbidden crystal-field transitions in $Fe^{2+}$ ions in M2 crystallographic positions of clinopyroxenes are responsible for a couple of weak bands near 0.505 μm and 0.550 μm (e.g., Burns et al., 1973; Platonov, 1976; Hazen et al., 1978; Matsyuk et al., 1985; Burns, 1993; Klima et al., 2006). If asteroid surface matter consists of a clinopyroxene and Fe orthopyroxene mixture then an additional weak band can originate at 0.525 μm due to the spin-forbidden crystal-field electronic *d-d*-transitions in $Fe^{2+}$ ions in M1 crystallographic sites of Fe orthopyroxenes. Moreover, a common absorption band at 0.49-0.55 μm (as any other one) could be widened because of disordering crystal structure at the process of space weathering of the asteroid surface (e.g., Charette et al., 1974; Gaffey et al., 1993). On the other hand, as shown in investigations of terrestrial analog/meteoritic samples (Rossman, 1975; Bakhtin 1985; Matsyuk et al., 1985; Khomenko and Platonov, 1987; Taran and Rossman, 2002), some intense absorption bands could arise at the same positions as spin-forbidden crystal-field electronic *d-d*-transitions in $Fe^{2+}$ and $Fe^{3+}$ ions. Probable mechanisms of such intensification are additional electronic transitions in magnetic exchange-coupled pairs (ECP) of neighboring cations, $Fe^{2+}$ - $Fe^{3+}$ (Mattson and Rossman, 1984; Matsyuk et al. 1985) or $Fe^{3+}$ - $Fe^{3+}$ (Rossman, 1975; Sherman, 1985; Sherman and Waite, 1985; Taran and Rossman, 2002) having common ligands. We do not exclude that these mechanisms could strengthen the discussed Pomona's absorption band at 0.49-0.55 μm. This assumption seems to be supported by the presence of additional absorption band in the reflectance spectra of Pomona at 0.73-0.77 μm (Fig. 1a). It could originate due to electronic $^6A_1 \rightarrow {}^4T_1$ transitions in $Fe^{3+}$ ions and located on the short-wavelength wing of the intense $Fe^{2+}$ absorption band (spin-allowed) of pyroxene-olivine mixture centered at 0.9-1.0 μm (Fig. 1a). Its parameters are similar to those of absorption bands registered in spectra of $Fe^{3+}$-bearing terrestrial and lunar pyroxenes (e.g., Mattson and Rossman, 1984; Bakhtin, 1985; Matsyuk et al., 1985; Khomenko, Platonov, 1987; Straub, Burns, 1990; Burns, 1993). Thus, the observed spectral features of Pomona are presumably the manifestations of predominant pyroxene – olivine mixture (e. g., Singer, 1981) and its various oxidations (e. g., Straub, Burns, 1990) in the surface matter.

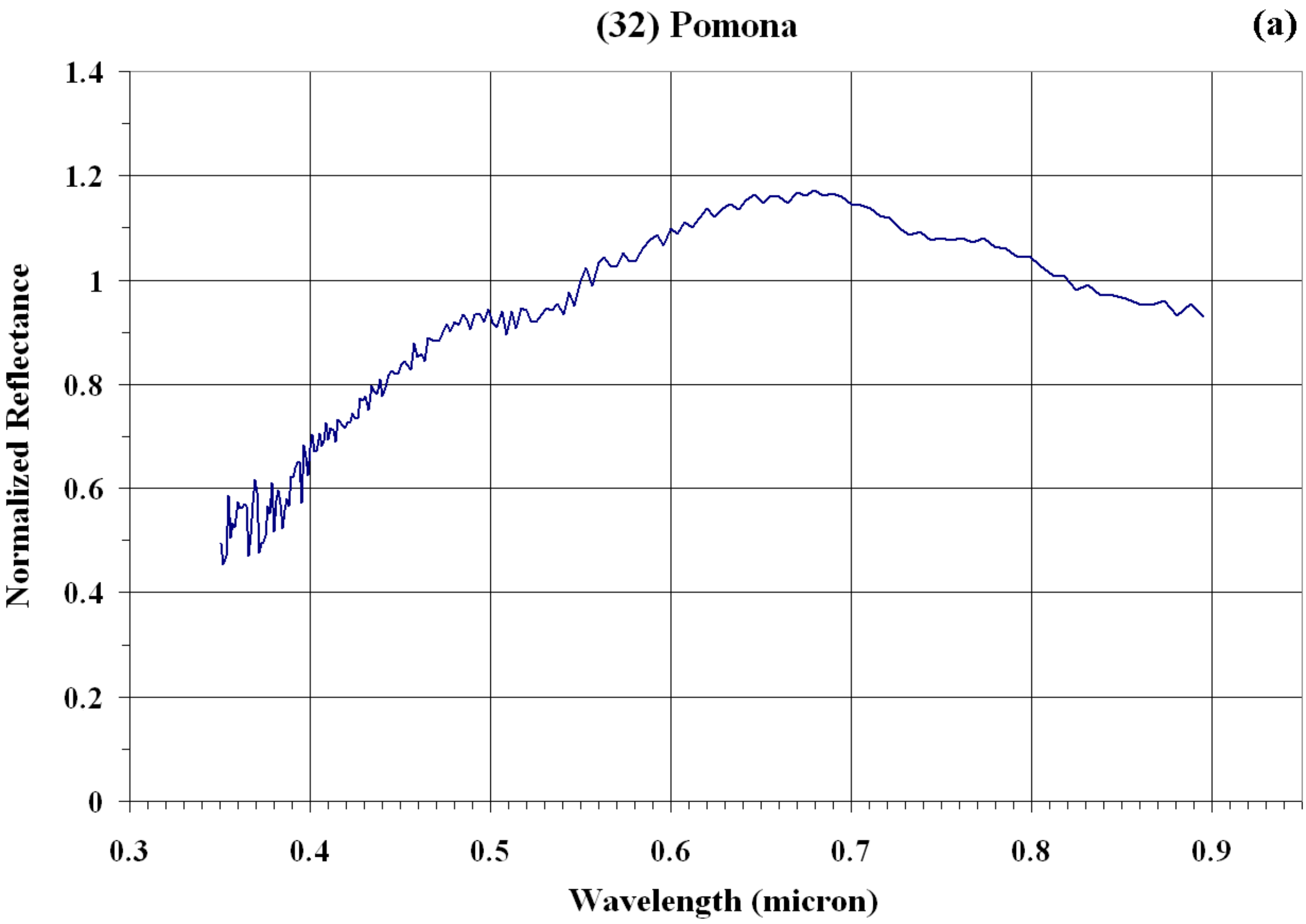

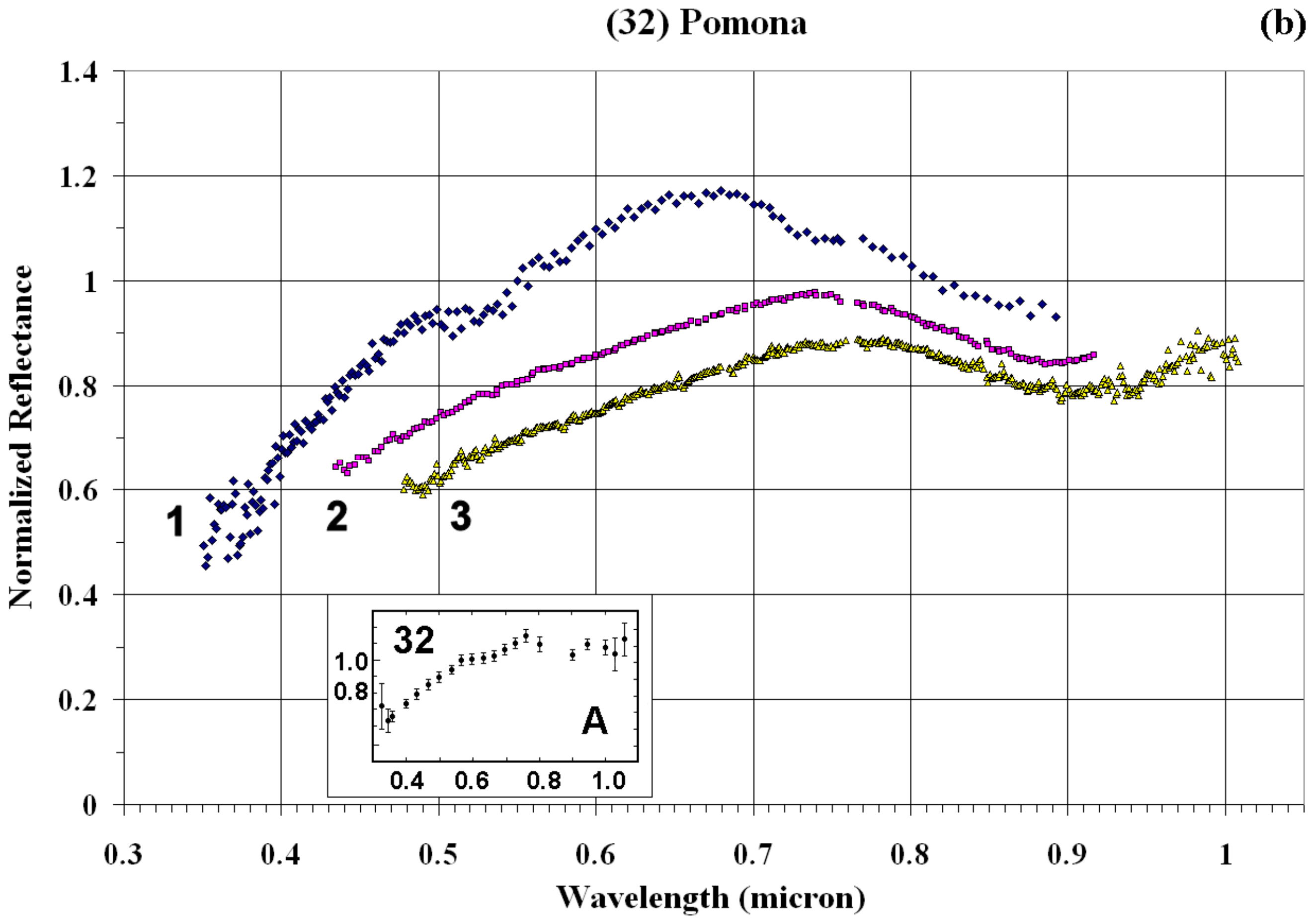

Figs. 1 (a, b). (a) Averaged and normalized at 0.55 μm reflectance spectrum of (32) Pomona. (b) Comparison of the same spectrum (1) with Pomona's spectra obtained in SMASSII (Bus and Binsel, 2003a) (2), in SMASS (Xu et al., 1995) (3), and by McCord and Chapman (1975a) (inset "A").

### 3.2. (145) Adeona, (704) Interamnia, and (779) Nina

(145) Adeona is a 126.13-km asteroid with geometric albedo of 0.06 (averaged WISE-data from Masiero et al., 2014) rotating with a period of $15.071^h$ (Harris et al., 2012). Its Tholen and SMASSII spectral types (Tholen, 1989; Bus and Binzel, 2002b) are C and Ch (”h” means that the asteroid is hydrated because of the presence of a 0.7 μm absorption band, according to Bus and Binzel, 2002b), respectively. The resulting averaged and normalized (at 0.55 μm) reflectance spectrum of Adeona is shown in Figure 2a. We have found three absorption bands in the spectrum. The strongest one is located between 0.48-0.85 μm. Judging by its parameters (Table 2) after removing the continuum slope (Fig. 5), the band probably arises due to electronic intervalence charge transfer (IVCT) transitions $Fe^{2+} \rightarrow Fe^{3+}$ in hydrated silicates, as follows from investigations of terrestrial samples (e. g., Platonov, 1976; Bakhtin, 1985; Burns, 1993). The second one, revealed itself between 0.40-0.47 μm (Fig. 2a), is probably due to electronic ECP-transitions in exchangeable $Fe^{3+}$ – $Fe^{3+}$ pairs ($^6A_1(^6S) \rightarrow {}^4T_2(^4G)$) in oxides, hydrated oxides and hydrated silicates (Sherman, 1985; Sherman and Waite, 1985; Bakhtin, 1985; Khomenko and Platonov 1987). The presence of a third band with relative intensity lower than ~5% is suspected at 0.36-0.39 μm (Fig. 2a and 7, Table 2), though it is comparable to the spectral noise in the range. Such a band might also be formed due to electronic ECP-transitions in $Fe^{3+}$ - $Fe^{3+}$ pairs ($^6A_1(^6S) \rightarrow {}^4E(^4D)$) (Sherman, 1985; Bakhtin, 1985; Khomenko and Platonov, 1987). We think that the detection of the two latter bands was possible thank to a high-altitude location of our observatory favorable for observations in the near-UV. We compared our spectra of Adeona with those available in the literature (Fig. 2b). Actually, the presence of absorption bands at 0.38 and 0.44 μm in reflectance spectra of the asteroid were not previously reported. This is likely because the published spectra have different wavelength ranges, in particular less extended to the blue. A weak 0.43-μm absorption band was detected in reflectance spectra of other eleven asteroids of primitive C, P, G classes by Vilas et al. (1993). A notable absorption band (~10%) centered near 0.68-0.70 μm presents in reflectance spectra of Adeona according to McCord and Chapman (1975b) (in Fig. 2b, inset "A") and Fornasier et al. (2014). Similar absorption features were found in reflectance spectra of many other asteroids (Vilas and Gaffey, 1989; Vilas et al., 1994; Bus and Binzel, 2002a; Fornasier et al., 2014). It is attributed to $Fe^{2+} \rightarrow Fe^{3+}$ IVCT-transitions in hydrated silicates (Vilas and Gaffey, 1989). Bus and Binzel (2002b) used the band in their classification system to delineate C and Ch (as well Cg and Cgh) type asteroids. However, we see that 0.7-μm absorption band in our reflectance spectrum of Adeona is several times more intense then that in data of McCord and Chapman (1975b) and Fornasier et al. (2014). There is also an unexplainable sharp rise of reflectivity in interval of 0.35-0.55 μm in our data of Adeona (Fig. 2b). We thoroughly checked our observational data but did not find any faults. We suppose that the difference is real and should be explained. We will try to do this by the end of the section.

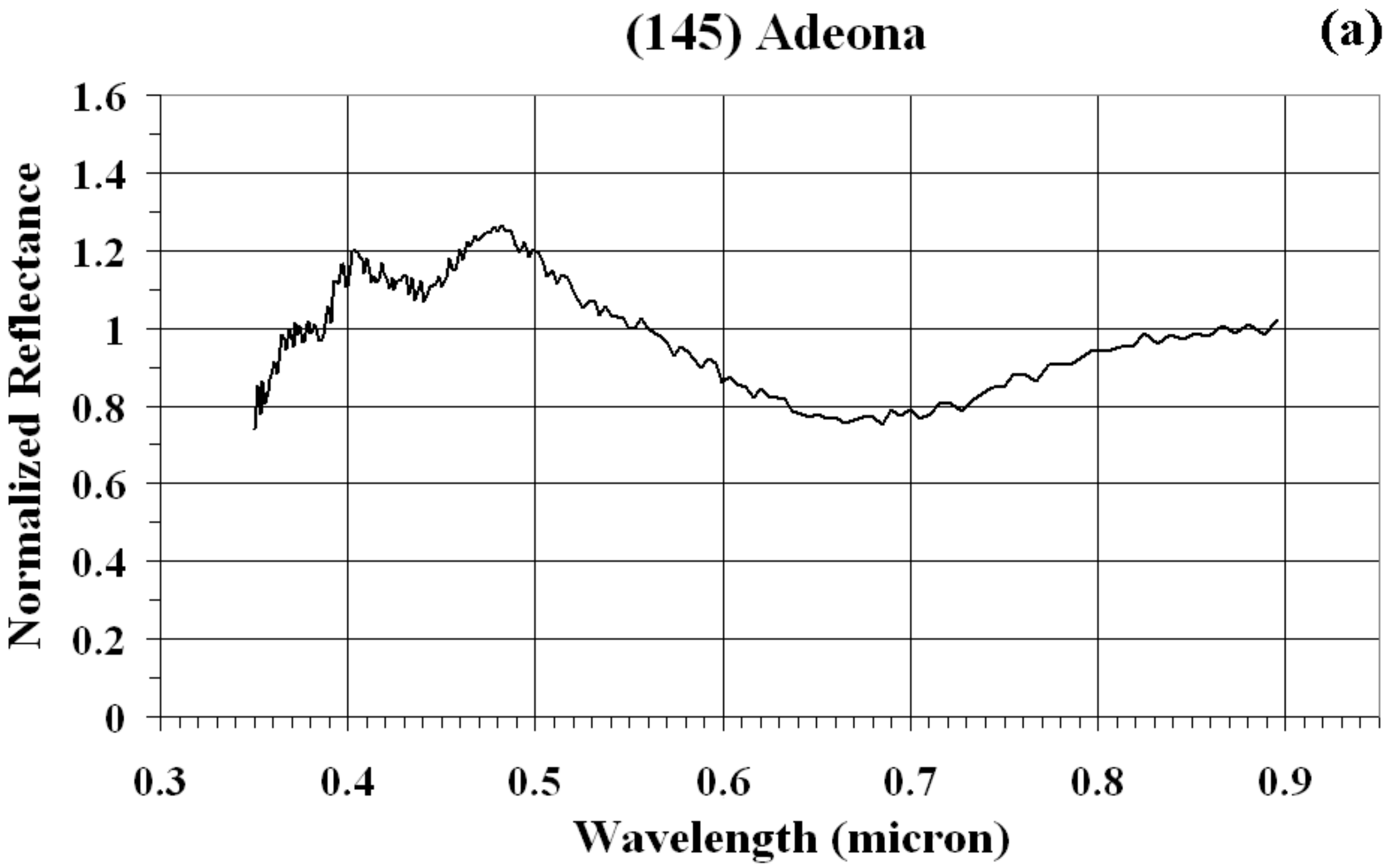

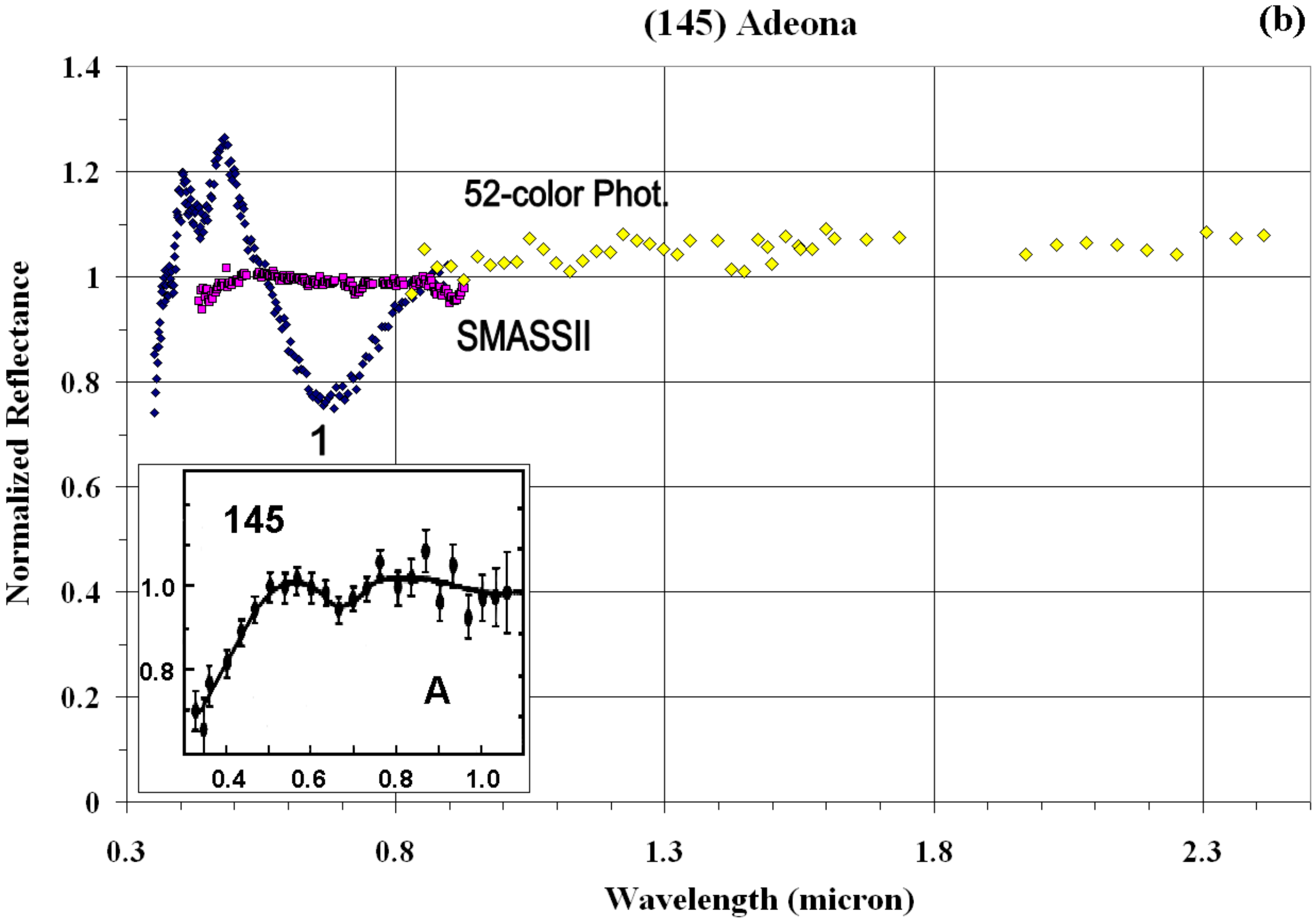

Figs. 2 (a, b). (a) Averaged and normalized at 0.55 μm reflectance spectrum of (145) Adeona. (b) Comparison of the same spectrum (1) with Adeona's spectra obtained in SMASSII (Bus and Binsel, 2003b), in 52-color photometric survey (Bell et al., 1995), and by McCord and Chapman (1975b) (inset "A").

Average diameter and geometric albedo of (704) Interamnia are 307.31 km and 0.08 according to WISE-data (Masiero et al., 2014). It rotates with a period of $8.727^h$ (Harris et al., 2012). The asteroid was previously classified as F-type body by (Tholen, 1989) and as B-type one by Bus and Binzel (2002b) characterizing by a wide absorption band beyond ~0.5 μm coupled with an overall negative gradient. In total, five spectra of Interamnia were obtained for ~3/4 hour corresponding to a change of its relative rotation phase ~0.08 (Table 1). We averaged these similar in shape spectra and showed as one resulting normalized (at 0.55 μm) spectrum in Figure 3a. It appears similar to that of Adeona, with virtually the same shape and relative intensity of features centered at approximately 0.38, 0.46, and 0.70 μm (see Figs 2a and 3a). Therefore, the same interpretation to them should be given. It should be also taken into account that slightly different averaging methods have been used for calculation of the mentioned reflectance spectra of the asteroid. While we and Fornasier et al. (2014) used the method of simple averaging, Bus and Binzel (2002a) applied the method of weighted averaging. In turn, Lazzaro et al. (2004) left a single asteroid spectrum (from several) with the highest S/N ratio. After the continuum slope removal, the absorption spectral features and their parameters are presented in Figs. 5-7 and Table 2. Again, we find that the majority of Interamnia's spectra available in the literature are less extended to the blue than ours (in particular, Bus and Binzel, 2002a; Lazzaro et al., 2004; Fornasier et al., 2014) except for that by Chapman et al. (1973) obtained in a wider spectral range. We compare some of them with ours in Figure 3b. As in the case of Adeona, our reflectance spectrum of Interamnia demonstrates a sharp rise of reflectivity at 0.4-0.7 μm clearly seen on a wider scale (Fig. 3a). We will return to this point in general, after consideration data on the next asteroid.

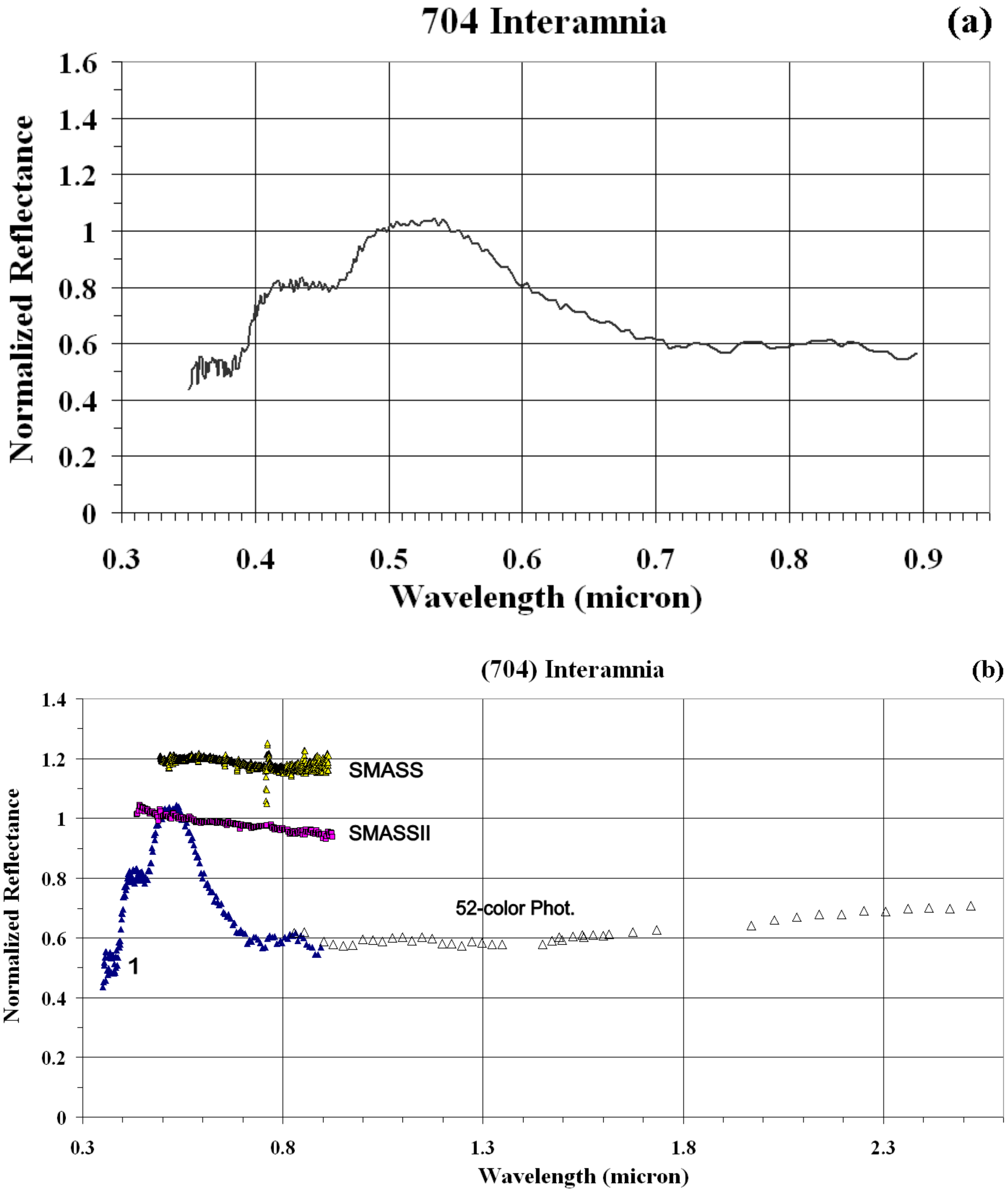

Figs. 3 (a, b). (a) Averaged and normalized at 0.55 μm reflectance spectrum of (704) Interamnia. (b) Comparison of the same spectrum (1) with Adeona's spectra obtained in SMASS (Lazzaro et al., 2006), in SMASSII (Bus and Binsel, 2003c), and in 52-color photometric survey (Bell et al., 1995).

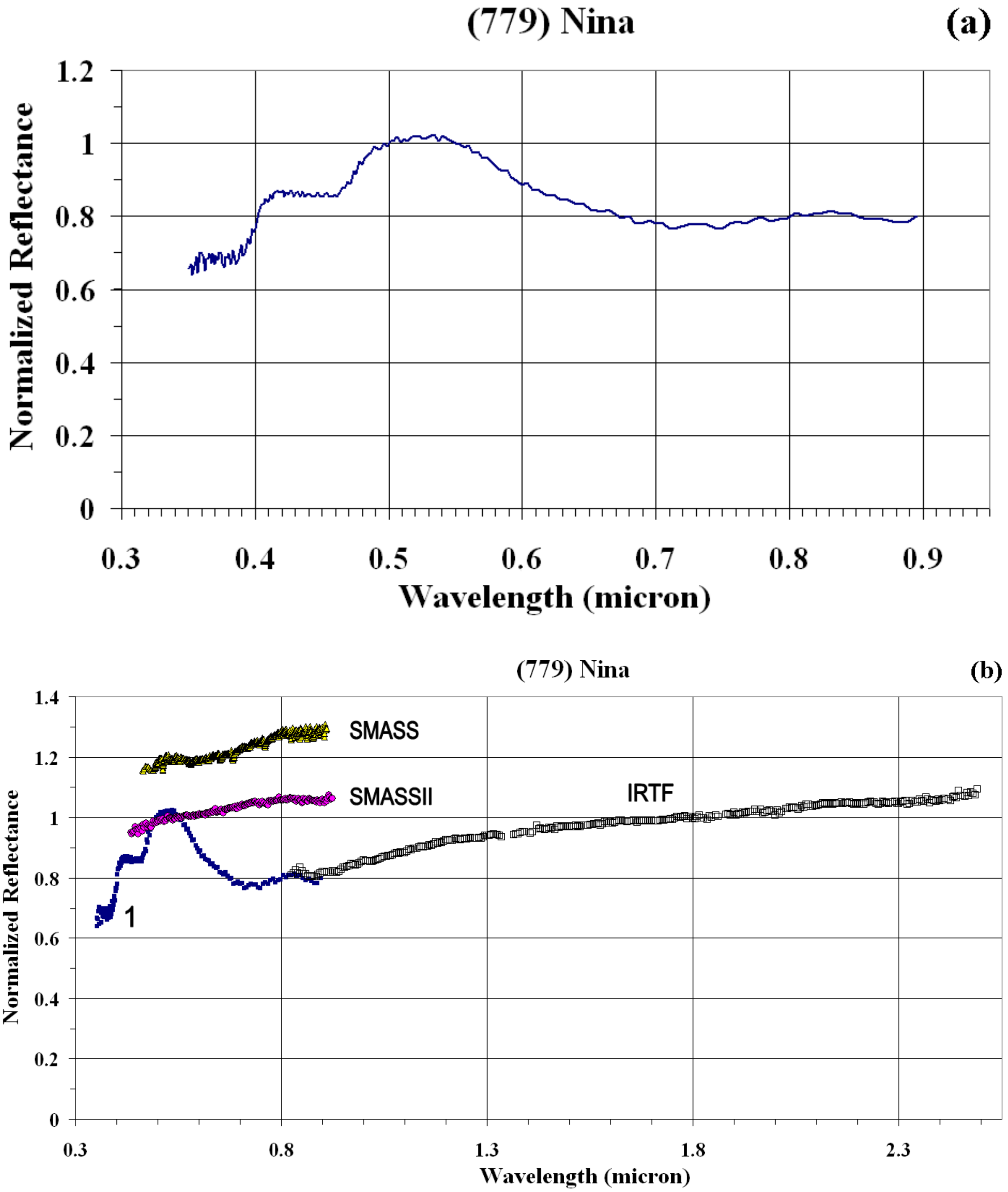

Figs. 4 (a, b). (a) Averaged and normalized at 0.55 μm reflectance spectrum of (779) Nina. (b) Comparison of the same spectrum (1) with Nina's spectra obtained in SMASS (Lazzaro et al., 2006), in SMASSII (Bus and Binsel, 2003d), and with IRTF (Ockert-Bell et al., 2011).

(779) Nina is a 80.57 km asteroid with geometric albedo of 0.16 according to WISE-data (Masiero et al., 2014) rotating with a period of $11.186^h$ (Harris et al., 2012). It was previously classified as M-type asteroid by Tholen (1989) and as X-type one according to SMASSII taxonomy (Bus and Binzel, 2002b).

In total, twelve observational spectra of Nina were obtained (Table 1). During the observations, Nina has rotated by ~0.5 of its period. One can see that, in fact, the presented reflectance spectrum of Nina (Fig. 4a) is similar to that of Interamnia in (Fig. 3a). There are virtually the same absorption bands in the spectra of Nina (located approximately at 0.39, 0.46 and 0.71 μm), Adeona, and Interamnia (see Fig. 4a and 5-7).

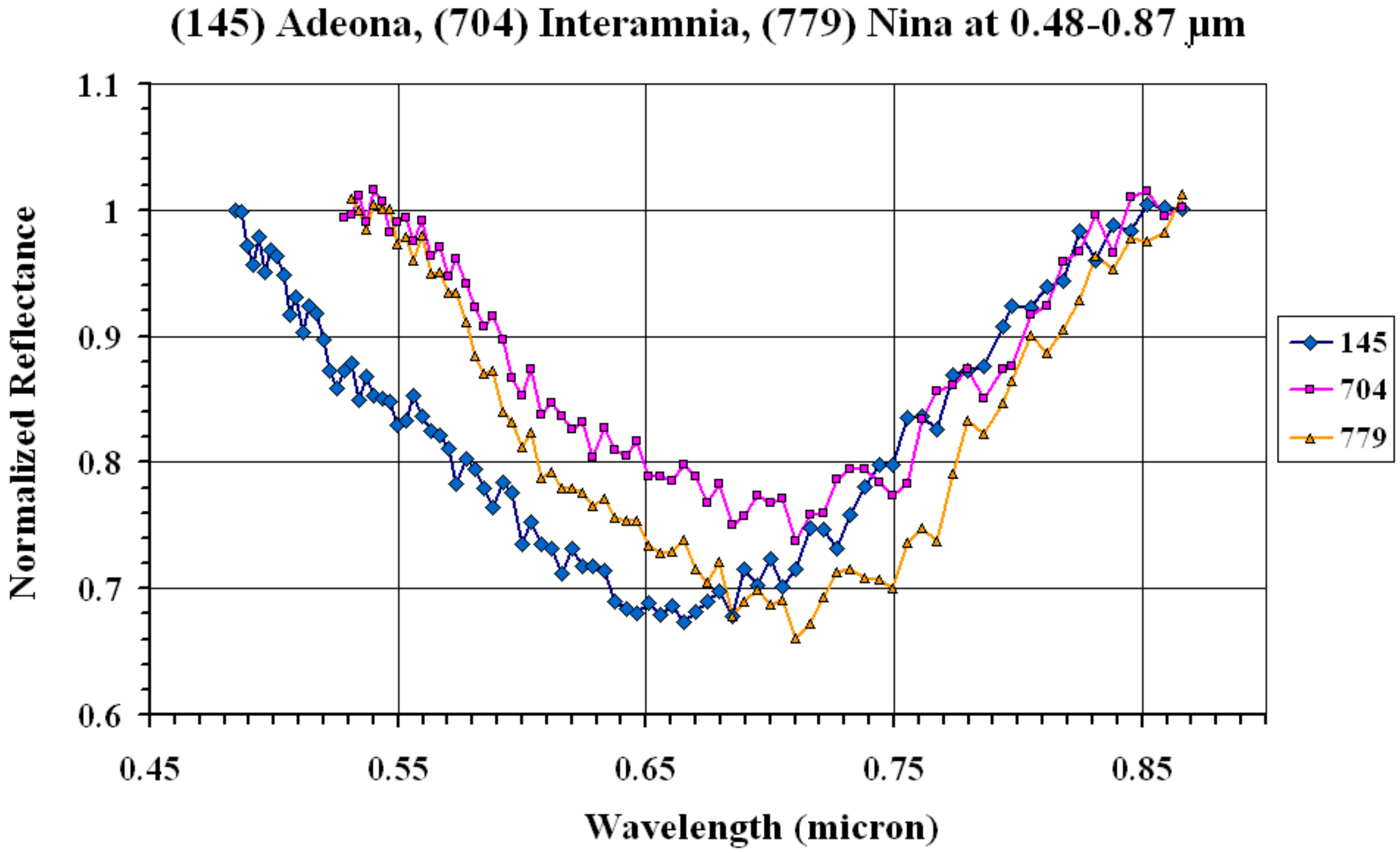

Fig. 5. Normalized absorption bands centered at ~0.50-0.71 μm in the reflectance spectra of (145) Adeona, (704) Interamnia, and (779) Nina after approximation and removal of the continuum.

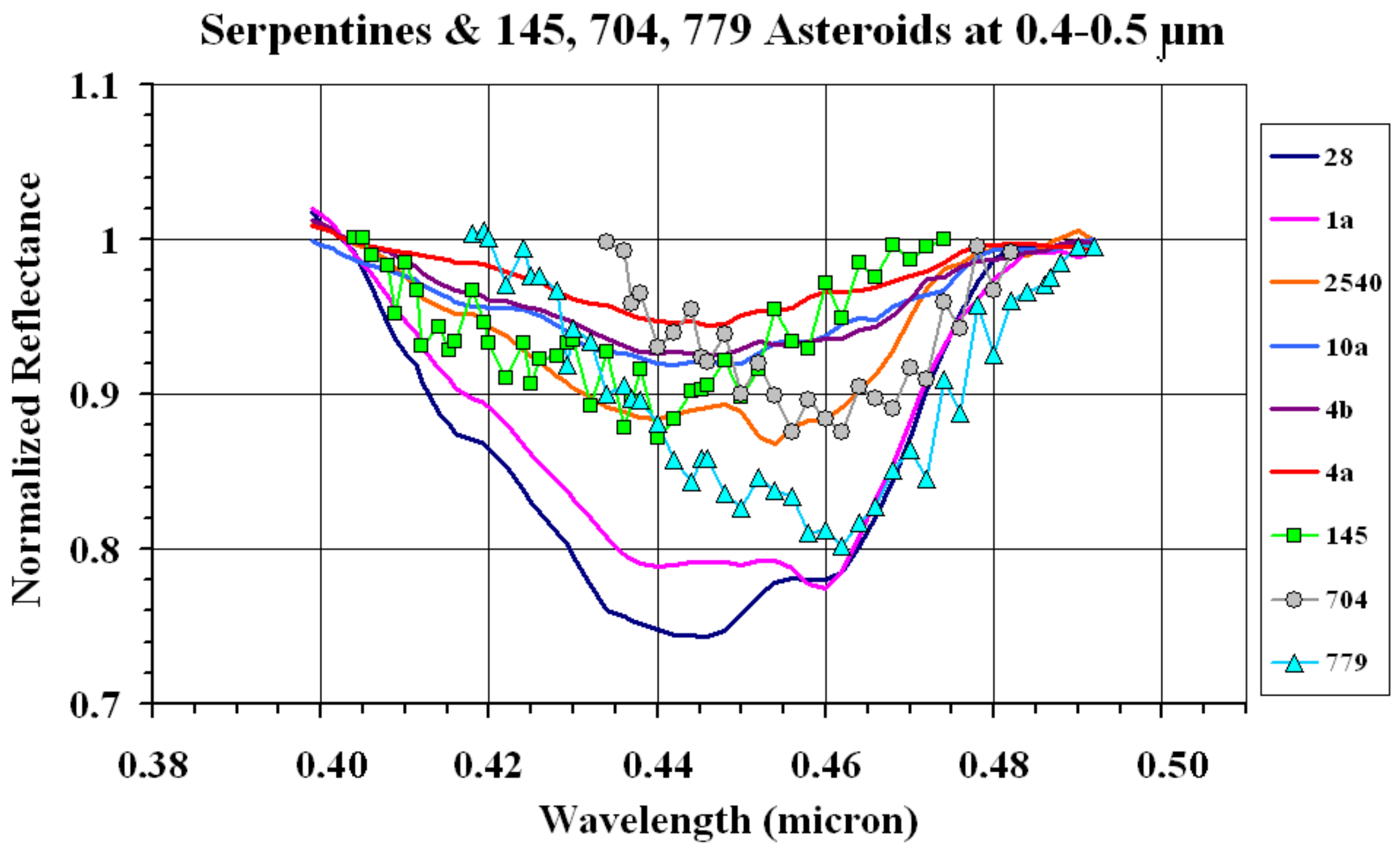

Fig. 6. Normalized absorption bands centered at 0.44-0.46 µm in the reflectance spectra of (145) Adeona, (704) Interamnia, and (779) Nina along with serpentine samples (numbered as 1a, 4a, 4b, 10a, 28, and 2540) after approximation and removal of the continuum.

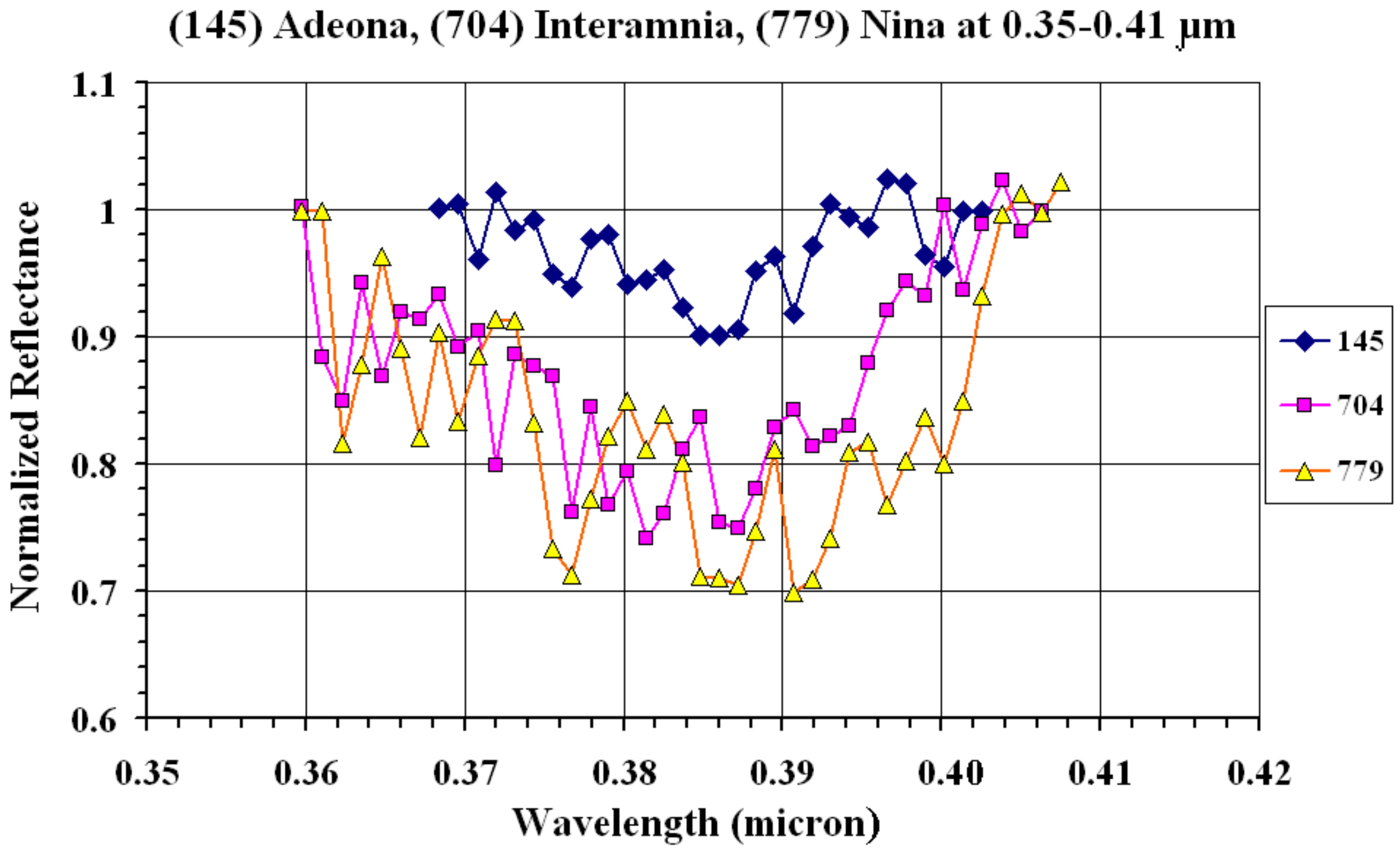

Fig. 7. Normalized absorption bands centered at 0.38-0.39 µm in the reflectance spectra of (145) Adeona, (704) Interamnia, and (779) Nina after approximation and removal of the continuum.

Table 2. Parameters of absorption bands detected in reflectance spectra of observed asteroids, carbonaceous chondrites and terrestrial serpentines.

| Spectral parameters / Object | Band center (μm) | Band width (μm) | Band relative intensity (%) | Interpretation | References |
|---|---|---|---|---|---|
| Asteroids: | | | | | |
| (32) Pomona | 0.53 | 0.49-0.56 | ~5-7% | spin-forbidden *d-d*-transitions in $Fe^{2+}$ strengthened by ECP-transitions in $Fe^{2+}$ - $Fe^{3+}$ pairs | [1-2, 7-9] |
| | 0.74 | 0.69-0.77 | ~3-5% | spin-forbidden *d-d*-transitions in $Fe^{3+}$ strengthened by ECP-transitions in $Fe^{3+}$ - $Fe^{3+}$ pairs | |
| | | | | spin-allowed *d-d*-transitions in $Fe^{2+}$ | [1-2, 7-9] |
| | ~0.90-1.0 | 0.68-? | ~27% | | [3, 6-8] |
| (145) Adeona | 0.38 | 0.37-0.39 | ~8%? | ECP- transitions in $Fe^{3+}$ - $Fe^{3+}$ pairs | [4-9] |
| | 0.44 | 0.41-0.47 | ~10%? | ECP- transitions in $Fe^{3+}$ - $Fe^{3+}$ pairs | [4-9] |
| | 0.67 | 0.49-0.85 | ~32%? | IVCT- transitions $Fe^{2+} \rightarrow Fe^{3+}$ | [3, 6-8] |
| (704) Interamnia | 0.38 | 0.36-0.40 | ~20%? | ECP- transitions in $Fe^{3+}$ - $Fe^{3+}$ pairs | [4-9] |
| | 0.46 | 0.43-0.48 | ~11%? | ECP- transitions in $Fe^{3+}$ - $Fe^{3+}$ pairs | [4-9] |
| | 0.70 | 0.55-0.85 | ~24%? | IVCT- transitions $Fe^{2+} \rightarrow Fe^{3+}$ | [3, 6-8] |
| (779) Nina | 0.39 | 0.36-0.41 | ~25% ? | ECP- transitions in $Fe^{3+}$ - $Fe^{3+}$ pairs | [4-9] |
| | 0.46 | 0.42-0.49 | ~19% ? | ECP- transitions in $Fe^{3+}$ - $Fe^{3+}$ pairs | [4-9] |
| | 0.71 | 0.55-0.85 | ~32%? | IVCT- transitions $Fe^{2+} \rightarrow Fe^{3+}$ | [3, 6-8] |
| (330825) 2008 XE3 | - | - | - | | |
| 2012 QG42 | 1.0? | 0.50-? | ~50% | spin-allowed *d-d*-transitions in $Fe^{2+}$ and/or IVCT- transitions $Fe^{2+} \rightarrow Fe^{3+}$ | [3, 6-8] |
| Carbonaceous chondrites: | | | | | |
| Orguel (CI) | 0.49 | 0.46-0.54 | ~5% | spin-forbidden *d-d*-transitions in $Fe^{2+}$ strengthened by ECP- transitions in $Fe^{2+}$ - $Fe^{3+}$ pairs | [1-2, 7-9] |
| | 1.0 | 0.6-? | ~10% | IVCT- transitions $Fe^{2+} \rightarrow Fe^{3+}$ | [3, 6-8] |
| Mighei (CM2) | 0.80 | 0.58-1.0 | ~5% | IVCT- transitions $Fe^{2+} \rightarrow Fe^{3+}$ | [3, 6-8] |
| Murchison (CM2) | 0.75 | 0.59-0.82 | ~12% | IVCT- transitions $Fe^{2+} \rightarrow Fe^{3+}$ | [3, 6-8] |
| | 0.90 | 0.82-1.0 | ~10% | IVCT- transitions $Fe^{2+} \rightarrow Fe^{3+}$ | [3, 6-8] |
| Boriskino (CM2) | 0.90 | 0.54-? | ~22% | IVCT- transitions $Fe^{2+} \rightarrow Fe^{3+}$ | [3, 6-8] |
| Terrestrial serpentines: | | | | | |
| 1a | 0.365 | 0.360-0.370 | ~1% | spin-forbidden *d-d*-transitions in $Fe^{3+}$ | [1-2, 7-9] |
| | 0.390 | 0.380-0.400 | ~2% | spin-forbidden *d-d*-transitions in $Fe^{3+}$ | [1-2, 7-9] |
| | 0.450 | 0.408-0.485 | ~23% | ECP- transitions in $Fe^{3+}$ - $Fe^{3+}$ pairs | [4-9] |
| | 0.750 | 0.555-1.0 | ~20% | IVCT- transitions $Fe^{2+} \rightarrow Fe^{3+}$ | [3, 6-8] |
| 4a | 0.390 | 0.370-0.400 | ~1% | spin-forbidden *d-d*-transitions in $Fe^{3+}$ | [1-2, 7-9] |
| | 0.450 | 0.408-0.480 | ~6% | ECP- transitions in $Fe^{3+}$ - $Fe^{3+}$ pairs | [4-9] |
| 4b | 0.450 | 0.408-0.485 | ~7% | ECP- transitions in $Fe^{3+}$ - $Fe^{3+}$ pairs | [4-9] |
| 10a | 0.390 | 0.370-0.400 | ~1% | spin-forbidden *d-d*-transitions in $Fe^{3+}$ | [1-2, 7-9] |
| | 0.443 | 0.400-0.485 | ~8% | ECP- transitions in $Fe^{3+}$ - $Fe^{3+}$ pairs | [4-9] |
| | 0.750 | 0.690-0.790 | ~3% | IVCT- transitions $Fe^{2+} \rightarrow Fe^{3+}$ | [3, 6-8] |
| 28 | 0.365 | 0.360-0.370 | ~1% | spin-forbidden *d-d*-transitions in $Fe^{3+}$ | [1-2, 7-9] |
| | 0.390 | 0.380-0.400 | ~2% | spin-forbidden *d-d*-transitions in $Fe^{3+}$ | [1-2, 7-9] |
| | 0.445 | 0.408-0.482 | ~25% | ECP- transitions in $Fe^{3+}$ - $Fe^{3+}$ pairs | [4-9] |
| 30 | 0.445 | 0.400-0.480 | Signs | ECP- transitions in $Fe^{3+}$ - $Fe^{3+}$ pairs | [4-9] |
| | 0.860 | 0.650- ? | ~20% | IVCT- transitions $Fe^{2+} \rightarrow Fe^{3+}$ | [3, 6-8] |
| 2540 | 0.365 | 0.360-0.370 | ~1% | spin-forbidden *d-d*-transitions in $Fe^{3+}$ | [1-2, 7-9] |
| | 0.390 | 0.380-0.400 | ~3% | spin-forbidden *d-d*-transitions in $Fe^{3+}$ | [1-2, 7-9] |
| | 0.445 | 0.408-0.480 | ~13% | ECP- transitions in $Fe^{3+}$ - $Fe^{3+}$ pairs | [4-9] |

After the continuum slope removal, the absorption spectral features of Nina and their parameters are presented in Figs. 5-7 and Table 2. Some differences of Nina's absorption features are their somewhat larger width and slightly longer wavelength positions as compared to those of Adeona and Interamnia. Therefore, the origin of these absorption bands in the spectra of Nina and of their respective counterparts in the spectra of Adeona and Interamnia are consistent. It should be mentioned that there are differences between the average geometric albedo of Adeona (0.043), Interamnia (0.08), and Nina (0.16) related probably to differences in the predominant content of their surface matter. Yet, radar albedo of Nina is typical for a primitive type asteroid (Shepard et al., 2010). The presence of the spectral features at 0.38 and 0.44 μm cannot be checked in spectra of Nina, obtained by Bus and Binzel (2002a) and Lazzaro et al. (2004), because they are less extended to the blue as compared with ours. There are signs of one feature centered at 0.6 μm in the asteroid spectrum obtained in SMASS by Lazzaro et al. (2004) (Figure 4b). Again, as seen from our reflectance spectrum of Nina presented along with IRTF and other data in the range up to 2.5 μm (Fig. 4b), there is a considerable growth of reflectivity of the asteroid between 0.4 and 0.7 μm.

Thus, a similar unexplainable effect found in reflectance spectra of Adeona, Interamnia, and Nina, three asteroids of primitive types, observed by us within a week (Table 1). We noted that Interamnia, and Nina in the time of observations were near their perihelion distances, when temperature on the asteroid surface reached a maximum near the subsolar point. At the same time, approaching to the Sun Adeona was close to its middle heliocentric distance. We suspected that the registered unusual short-wavelength raise of asteroid reflectance may be result of a sublimation activity of the bodies' surface matter including ices. A cloud or coma of sublimed and frozen particles surrounding an asteroid could produce a considerable scattering of light reflected from the asteroid. Given spectrally neutral refractive index of the particles in the used range, the wavelength position of a "hump" created by scattering of reflected light in asteroid reflectance spectrum is likely determined by predominant particle sizes in the coma. As seen in figs. 2a, 3a, and 4a, it takes place around ~0.4-0.6 μm.

The surface temperatures of main-belt asteroids are estimated by direct measurements and modeling. For instance, the Rosetta/VIRTIS measurements showed that the surface temperature on (21) Lutetia varies between 170 and 245 K (Coradini et al., 2011). Among volatiles abundant near planetary surfaces (e. g., Dodson-Robison et al., 2009), the triple point in the phase diagram for $H_2O$ and $CO_2$ (and their mixtures) is at the closest position to this temperature range. However, the point of $CO_2$ is shifted downward by ~50 degrees (e. g., Longhi, 2005). Therefore, one could expect that it is $H_2O$ that predominantly survives under the surface of primitive-type asteroids since their formation or since previous falls of smaller ice bodies on the asteroids (e. g., Fanale and Salvail, 1989; Jewitt and Guilbert-Lepoutre, 2012). Moreover, parent bodies of C and close type asteroids should include a considerable ice component because of heliocentric zoning of the asteroid belt by $^{26}Al$ heating (Grimm and McSween, 1993). In turn, intensive sublimation of fresh $H_2O$ ice excavated on a primitive asteroid at recent impact(s) may create a coma of icy particles around the body at the highest surface temperatures near its shortest heliocentric distances.

To verify our assumption, we estimate the subsolar temperatures on the observed asteroids. As is known, most celestial objects (including planets) have approximately blackbody spectra and effective temperatures described by the Stefan-Boltzmann law. In addition, we relate the received solar electro-magnetic energy by the unit area at an arbitrary top level of terrestrial atmosphere at the subsolar point (that is the solar constant, $C_e = L_{sun}/4\pi\ r_e^2$ =1367 W·m$^{-2}$, in which $L_{sun}$ is the luminosity of the Sun, $r_e$ is Earth's heliocentric distance) and the radiative energy received by the unit area normal to the direction to the Sun at heliocentric distance $r$ (in AU): $E = C_e \cdot 4\pi\ r_e^2/4\pi \cdot r^2$. Accordingly, we calculate the effective temperature in the subsolar point of an atmosphereless solid body taking into account the Stefan-Boltzmann constant $\sigma$ = 5,6704·10$^{-8}$ J· s$^{-1}$· m$^{-2}$ ·T$^{-4}$ (or in W· m$^{-2}$ ·T$^{-4}$) and $r_e$ = 1 AU:

$$T_{ss} = ((1-p_v)\cdot C_e \cdot 4\pi\ r_e^2/4\pi \cdot \sigma \cdot r^2)^{1/4} = ((1-p_v)\cdot C_e \cdot\ r_e^2/\ \sigma \cdot r^2)^{1/4} = ((1-p_v)\cdot 241.08\cdot 10^8 \cdot\ K^4\ /\ r^2)^{1/4} \text{ or}$$

$$T_{ss} = 394K\cdot\ ((1-p_v)\ /\ r^2)^{1/4} \qquad (1)$$

Calculations by the formula give next subsolar temperatures on the asteroids at the moments of observations: $T_{ss,\ o}$ = 236.5 K (Adeona), $T_{ss,\ o}$ = 238.6 K (Interamnia), and $T_{ss,\ o}$ = 257.3 K (Nina). As it turns out, the Adeona's and Interamnia's temperatures are very similar. All the temperature values are in the temperature range of the most intense sublimation of water ice: within several tens of degrees below 273 K.

Our consideration of Adeona's, Interamnia's, and Nina's reflectance spectra (Figs 2-4), their interpretation and comparison with those of possible analog samples (Figs 10 and 11) show that scattering of reflected light in hypothetical asteroid coma of sublimed ice particles does not change position of intrinsic asteroid absorption bands. However, it may change their intensity, especially for 0.7-μm band which minimum almost coincides with the change of continuum gradient sign. For this reason, we have queried all the estimated values of relative intensity of the found absorption features (Table 2).

### 3.3. (330825) 2008 XE3 and 2012 QG42

Two NEAs were observed by us, as well. (330825) 2008 XE3, a member of Amor group, and 2012 QG42, PHA and a member of Apollo group (http://ssd.jpl.nasa.gov/sbdb.cgi#top), have rotational periods of 4.409$^h$ (Warner, 2013) and 24.22$^h$ (Warner et al., 2013), respectively. The obtained spectra of 330825 were registered for about 2.5 hours that corresponds to change in its rotational phase by ≈ 210°.

Normalized at 0.55 μm reflectance spectra of the asteroid are shown in Figure 8. Spectrum 2 is arbitrarily displaced relative to spectrum 1 to facilitate their comparison. Spectra 1 and 2 are the averages of three and two consecutive reflectance spectra, respectively. Negligible changes of the asteroid spectra with rotation are indications of a relatively homogenous content of the surface matter. Based on the overall shape of the spectra with a minor positive gradient, 330825 may be ascribed to taxonomic type S or C as follows from spectral features of other asteroids (Tholen and Barucci, 1989; Bus and Binzel, 2002a, b). Moreover, absence of typical for S-type asteroids olivine-pyroxene band at 0.9 μm in its reflectance spectrum makes 330825 similar to a C-type body. However, absence of 0.9-μm band could be result of existence of some darkening components (e. g., carbonaceous compounds or magnetite) in the surface matter of 330825. Thus, we are unable unambiguously to determine type of 330825 and, hence, its predominant mineralogy from our data, and state only that it may be C(S). Obviously, it should be additionally investigated.

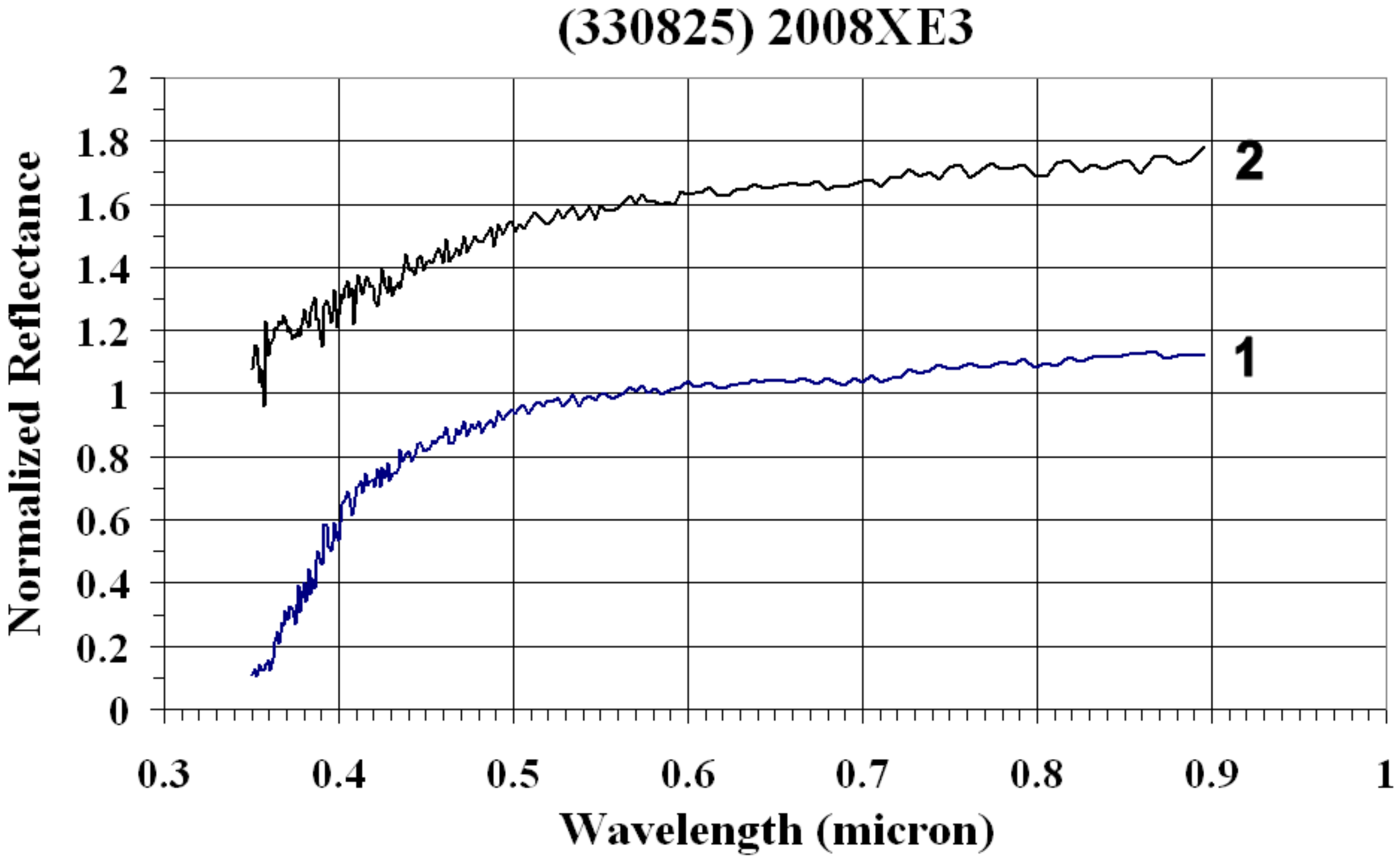

Fig. 8. Reflectance spectra of asteroid 330825 (2008 XE3) normalized at 0.55 μm, presented in the chronological order, and shifted arbitrarily for clarity. Spectra 1 and 2 are the averages of three and two similar in shape consecutive spectra, respectively.

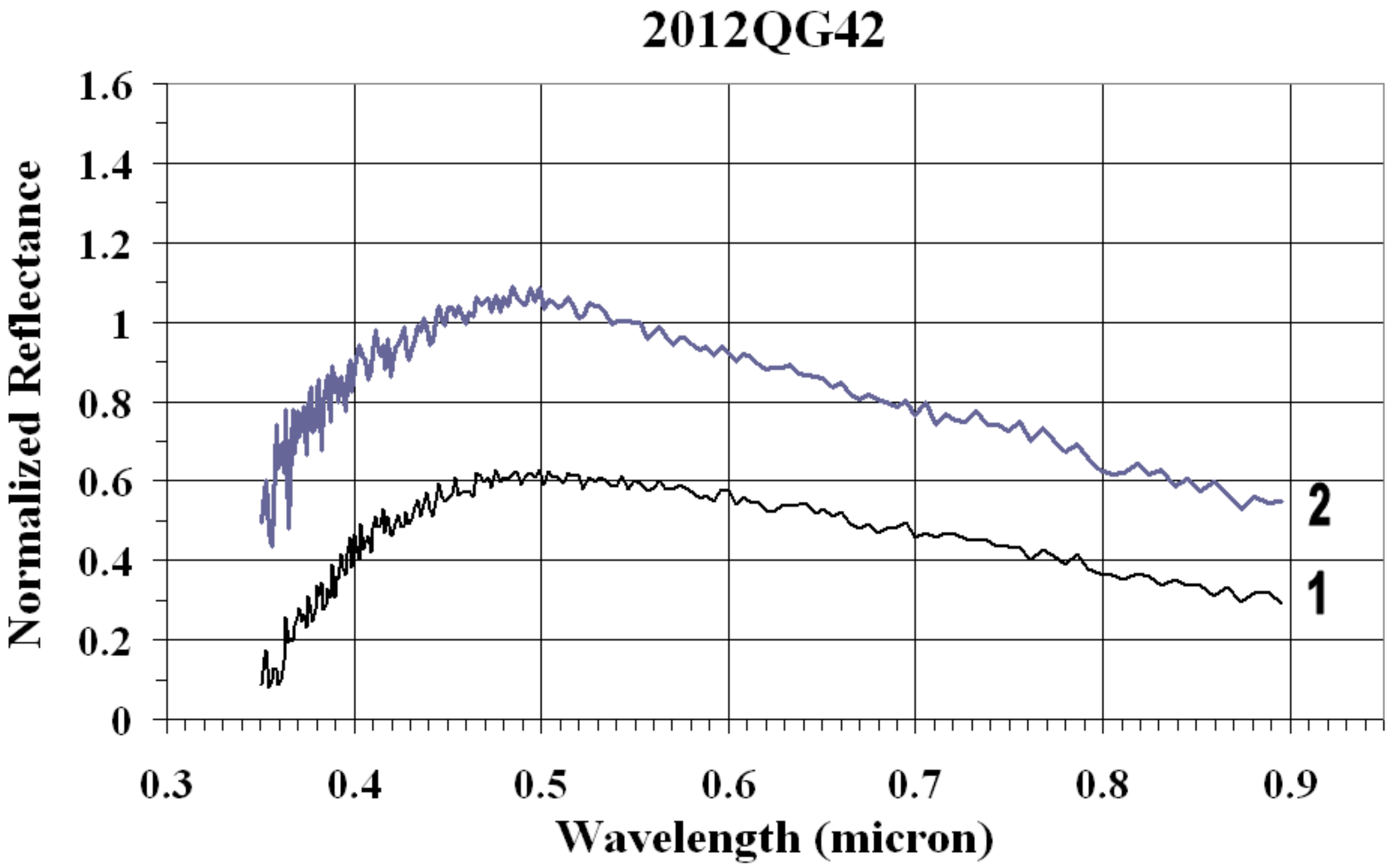

Fig. 9. Reflectance spectra of asteroid 2012 QG42 normalized at 0.55 μm, presented in the chronological order, and shifted arbitrarily for clarity. Spectrum 1 is the average of three similar in shape consecutive reflectance spectra, respectively. Reflectance spectrum 2 is obtained from a single observed spectrum of the asteroid.

Observational spectra of 2012 QG42 were registered for ≈ 2 hours, which corresponds to only ≈ 0.09 portion of its rotation period. Asteroid reflectance spectra normalized to 1 at 0.55 μm are shown in chronological order in Figure 9. The spectra are arbitrarily displaced for clarity of the plot. Spectrum 1 of 2012 QG42 is an average of three similar in shape consecutive reflectance spectra. Spectrum 2 is a single reflectance spectrum. It was separated from the averaged subset of the first three spectra (Table 1) to show their similarity in common. A maximum in the spectra of 2012 QG42 at 0.50 μm (Fig. 9) and possible extended short-wavelength wing of an absorption band at 0.9-1.0 μm makes them similar to spectra of either S-type asteroids or B-type ones having a negative gradient of reflectance spectra in the visible to near-IR range (Tholen and Barucci, 1989; Bus and Binzel, 2002b). Thus, we assume that it may be of S or B-type. The surface matter of 2012 QG42 could include not only high-temperature minerals (for instance, pyroxene, olivine, etc.) but also hydrated silicates. To distinguish exactly the type of 2012 QG42, it needs additional investigation of the asteroid in a wider spectral range or by other methods. At the same time, previous publications devoted to (330825) 2008 XE3 and 2012 QG42 stated explicitly that both bodies are of S-type (Warner et al., 2013; Taradii et al., 2013).

An important point should be discussed here. As seen from Figs. 8 and 9, reflectance spectra of both NEAs, (330825) 2008 XE3 and 2012 QG42, are featureless compared to the considered main belt asteroids. This suggests that the lack of subtle absorption features in reflectance spectra of near-Earth asteroids may be results of common environmental or space weathering factors. The most significant of them are higher surface temperatures and more strong darkening action of the solar wind and UV-radiation. As experimental modeling showed, the consequences of the latter are surface darkening and reddening of silicate materials from near-UV to near-IR ranges (e. g., Loeffler, et al., 2009; Kaˇnuchová et al., 2010). Formation of $H_2O$ film of amorphous ice is possible directly from the vapor phase onto the surface particles of an asteroid at temperatures below 130 K (Hobbs, 1974). On the contrary, at 1 AU from the Sun, typical temperatures at the subsolar point reach about 400 K, whereas many of the known NEAs have perihelion distances below 0.3 AU where the surface temperature can rise above 800 K (Delbo' et al., 2009). Thus, the sublimation rate of absorbed or buried near the surface water ice, if present, should considerably increase while the bodies are approaching the Sun. Existence of the surface interstitial ice or bound water on different type asteroids was proved by observations of its characteristic absorption band at 3.0 micron (Lebofsky et al., 1981; Jones et al., 1990; Rivkin et al., 1995, 2000). Also, as shown by modeling and calculations (e. g., Zolensky et al., 1989; Shock and Schulte, 1990), synthesis of water-soluble organic compounds was possible in the process of aqueous alteration of insoluble organics in the carbonaceous meteorite parent bodies. Thus, the main result of considered process of water ice sublimation on the asteroid surface should be creation of dark or insufficiently transparent organic films and/or blebs on the surface silicate particles. In turn, the dark films must preclude the formation of diffuse component in the reflected light containing information on the surface matter content. On the other hand, the presence of relatively transparent frozen volatiles in between the surface particles on the main-belt asteroids could make absorption bands in their reflectance spectra more strong. Another important effect of considerable temperature fluctuations on the surface of NEAs is mineral decomposition into silicate particles themselves followed by the disappearance of their spectral features. Then, it could explain the absence of weaker absorption features and even some changing the overall shape of reflectance spectra of near-Earth asteroids. For these reasons, taxonomic classification of the bodies is complicated.

## 4. Laboratory investigations of analog samples

We performed a laboratory study of some proper analog samples to facilitate interpretation of the obtained reflectance spectra of asteroids. Taking into account taxonomic S-type of (32) Pomona and its supposed pyroxene-olivine content confirmed by our measurements, we rely on previous results of other authors. Actually, reflectance spectra of pyroxenes, olivines and their mixtures are well documented (e. g., Adams, 1974; Singer, 1981; Cloutis et al., 1986; Khomenko and Platonov, 1987; Burns, 1993). On the other hand, it should be recognized that knowledge of spectral characteristics of analog samples for primitive type asteroids remains insufficient. Investigations of the asteroids themselves and the most primitive meteorites, as their possible fragments, showed that best spectral analogs of the bodies are carbonaceous chondrites, phyllosilicates (serpentines, chlorites, etc.), oxides and hydroxides (e. g., Gaffey et al., 1989; Calvin and King, 1997; Hiroi and Zolensky, 1999; Cloutis et al., 2011a, b). Carbonaceous chondrites are rare for their instability at falls in terrestrial atmosphere (shocking and heating) and subsequent rapid weathering (e. g., Dodd, 1981). For these reasons, such type of matter is incompletely represented in meteoritic collections. Taking also into account known taxonomic types of (145) Adeona, (704) Interamnia, and (779) Nina as primitive or close to that, we have decided to spare a special attention to CI-CM carbonaceous chondrites and serpentines as a predominant component of the meteorites and possibly surface matter of the asteroids. Thus, we performed a study of some useful CI-CM carbonaceous chondrite and phyllosilicate analog samples (Table 3.)

Table 3. Description of the used serpentine samples.

| Sample | Place of origin | The main constituents (in order of abundance) |
|---|---|---|
| 1a | the North Urals, Russia | chrysotile and lizardite |
| 4a | Pobuzh'e, Ukraine | lizardite, chrysotile and calcite |
| 4b | Pobuzh'e, Ukraine | lizardite, chrysotile and calcite |
| 10a | Bazhenovskoe deposit, Urals, Russia | chrysotile and lizardite |
| 28 | Bazhenovskoe deposit, Urals, Russia | lizardite and chrysotile |
| 30 | Idzhim deposit, Sayans, Russia | brucite, chrysotile and lizardite |
| 2540 | Nizin-Kol river, Kuban', Russia | chrysotile and lizardite |

**4.1. Carbonaceous chondrites.** We undertook reflectance measurements of accessible samples of carbonaceous chondrites. Four samples of known carbonaceous chondrites, Orguel (CI), Mighei (CM2), Murchison (CM2), and Boriskino (CM2), were taken from the meteoritic collection of Vernadsky Institute of Geochemistry and Analytical Chemistry (RAS). Reflectance spectra of the ground meteoritic samples (≤0.25 mm grain size) were measured in the 0.35-1.00 μm range with a single-beam spectrophotometer based on a SpectraPro-275 triple-grating monochromator controlled by a PC. The angles of incident and reflected light beam and its diameter were 45°, 0°, and 5 mm, respectively.

Compressed powder of MgO was used as a reflectance standard. The root mean square relative error for the reflectance spectra does not exceed 0.5–1.0% in the visible and increases gradually to 1–2% at the red end of the operational spectral range (Busarev and Taran, 2002).

As seen from the measured normalized reflectance spectra, carbonaceous chondrites of CI and CM groups (Fig. 10), have most prominent absorption bands at 0.45-0.57 and 0.80 or 0.75 and 0.90 μm (Table 2). The spectral features of CI-CM carbonaceous chondrites are in agreement with those found in previous works (e. g., Calvin and King, 1997; Cloutis et al., 2011a, b).

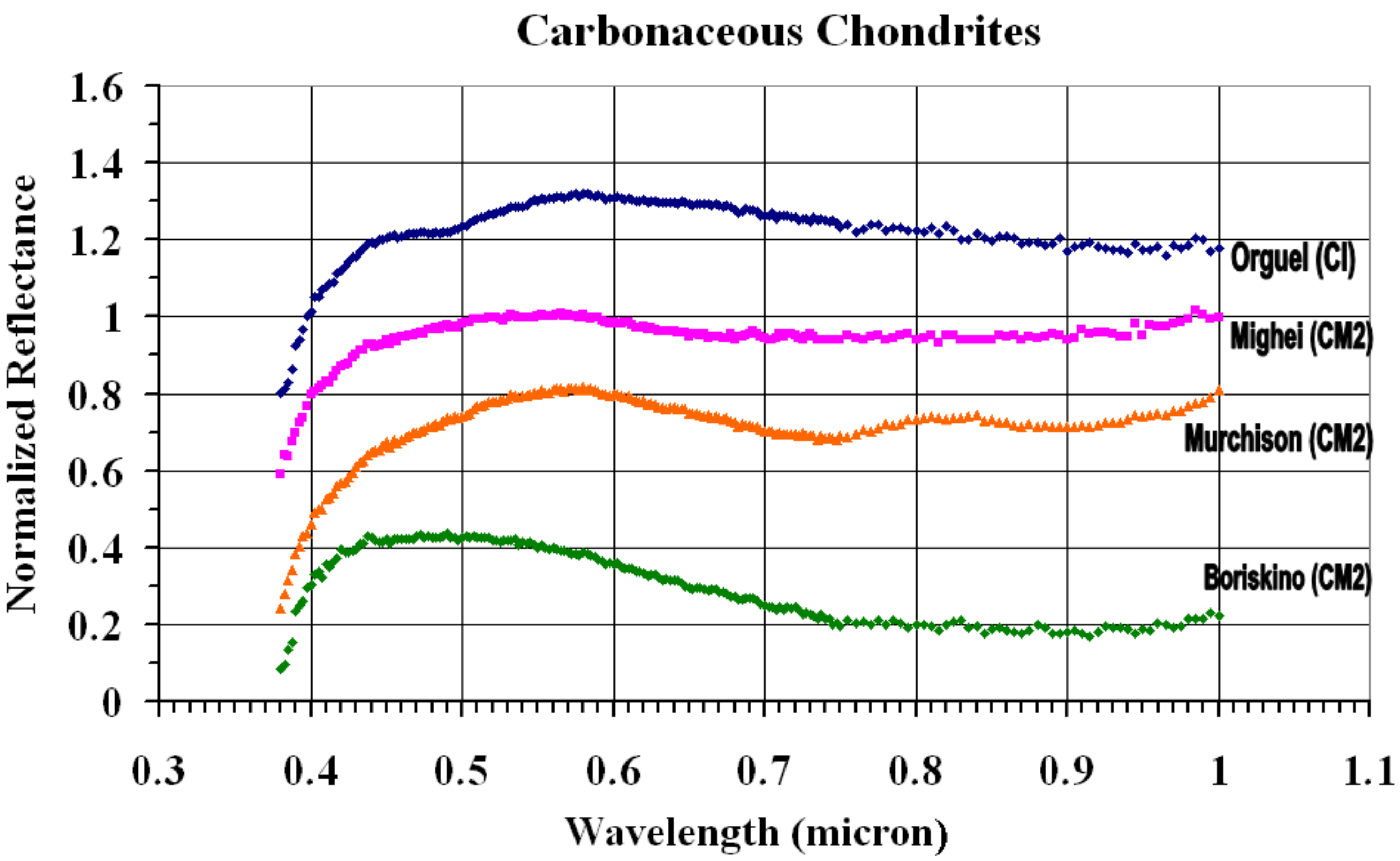

Fig. 10. Normalized at 0.55 μm and shifted reflectance spectra of powders (particle sizes ≤ 0.25 mm) of carbonaceous chondrites, Orguel (CI), Mighei (CM2), Murchison (CM2), and Boriskino (CM2) (particle sizes ≤ 0.25 mm), taken from the meteoritic collection of Vernadsky Institute of Geochemistry and Analytical Chemistry of RAS.

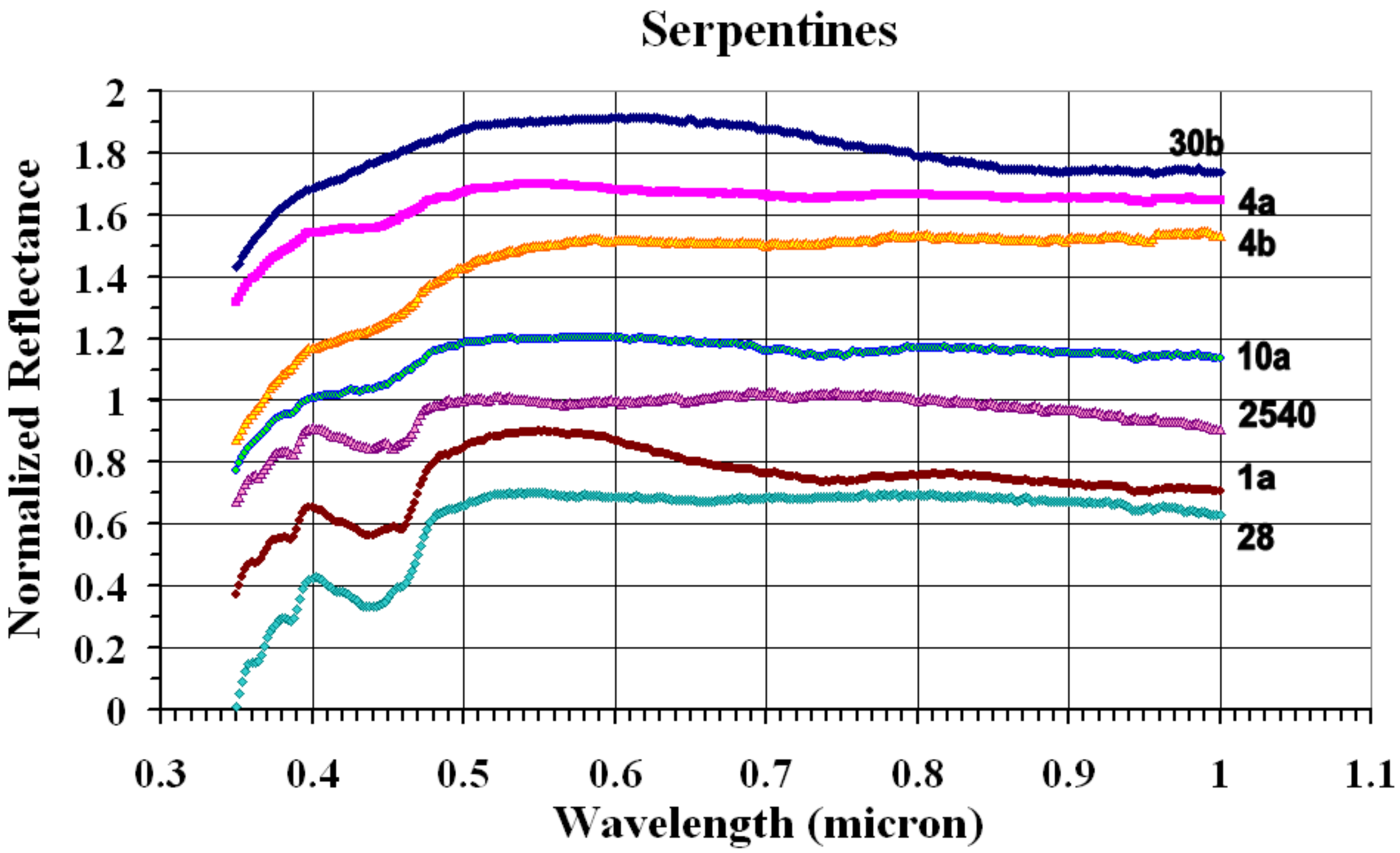

Fig. 11. Normalized at 0.55 µm and shifted reflectance spectra of powdered terrestrial serpentines (numbers 1a, 4a, 4b, 10a, 28, 30, and 2540) (particle sizes ≤ 0.15 mm). The spectra are disposed from top to bottom in the order of intensity increase of an absorption band at 0.40-0.48 µm.

**4.2. Serpentines.** Seven low-iron Mg serpentine samples were also taken as possible analogs for primitive type asteroids. It should be noted that such a type of serpentine is typical for an early stage of serpentinization (e. g., Deer et al., 1963; Barber, 1981; Brearley, 2006). The serpentine samples were picked out from original ones found in different geological formations in Russia and Ukraine (Table 3). Reflectance spectra of the ground serpentine samples (particle sizes ≤ 0.15 mm) were measured with the above described single-beam spectrophotometer. The obtained reflectance spectra of the samples are shown in Figure 11. There are absorption bands at 0.40-0.50 and 0.80 or 0.75 and 0.90 µm in the spectra (Fig. 11). Positions of the features agree with those found in earlier works (e. g., King and Clark, 1989; Calvin and King, 1997; Hiroi and Zolensky, 1999). We noted that the most prominent one centered at 0.435-0.445 µm has different relative intensity (up to ~25%) in the spectra (Table 2). Excluding our publication (Busarev et al., 2004), a considerable intensity of the absorption band in reflectance spectra of serpentine samples was not previously reported. There is the only published information on detection of an intense absorption band at 0.44 µm in $Fe^{3+}$- content ($Fe_2O_3 \approx 0.5$ wt. %) yellow corundum ($\alpha \approx Al_2O_3$) (Taran et al., 1993).

Table 4. Elemental analyses for serpentine samples.

| Sample | 1a | 4a | 4b | 10a | 28 | 30b | 2540 |
| --- | --- | --- | --- | --- | --- | --- | --- |
| Oxides (wt. %) | | | | | | | |
| $FeO/Fe_2O_3$ | 2.11 | 1.35 | 1.80 | 1.91 | 2.62 | 2.19 | 2.36 |
| MgO | 41.98 | 38.56 | 51.61 | 38.43 | 38.50 | 34.65 | 38.17 |
| $Al_2O_3$ | 0.67 | - | - | 0.76 | 1.20 | 0.08 | 1.25 |
| $SiO_2$ | 41.33 | 39.75 | 36.77 | 43.72 | 43.30 | 39.51 | 42.50 |
| $Cr_2O_3$ | - | - | 0.06 | 0.60 | 1.07 | 0.11 | 1.34 |
| CaO | 0.02 | 0.06 | 0.04 | 0.09 | 0.01 | 0.06 | 0.02 |
| $K_2O$ | 0.01 | 0.01 | - | - | - | 0.02 | 0.08 |
| $Na_2O$ | - | 0.06 | - | - | 0.17 | 0.17 | 0.03 |
| MnO | 0.14 | - | 0.01 | 0.08 | 0.07 | 0.01 | - |
| $P_2O_5$ | 0.11 | 0.11 | - | 0.02 | 0.07 | 0.01 | 0.17 |
| $TiO_2$ | 0.04 | - | - | - | - | - | - |
| NiO | - | - | - | - | - | - | - |
| Sum | 86,40 | 79,90 | 79,9 | 85,60 | 87.00 | 76.80 | 85.90 |

Remarks: The sum of oxides is calculated excluding volatile loses.

Table 5. The relative intensity of the Mössbauer partial spectra.

| Sample | Paramagnetic partial spectra | | | | Magnetite ($Fe_3O_4$) |
|---|---|---|---|---|---|
| | I($Fe^{3+}$), % | | I($Fe^{2+}$), % | | I($Fe^{3+}$), % |
| | Td | Oh | Oh | Td | Td, Oh |
| 1a | 30.0(1.9) | 47.3(2.4) | 22.7(1.4) | - | - |
| 4a | 33.0(6.3) | 27.9(6.9) | 39.2(1.6) | - | - |
| 4b | 33.1(4.0) | 15.3(3.0) | 51.5(2.8) | - | - |
| 10a | 4.8(1.4) | 36.9(4.6) | 25.2(2.9) | - | 21.2(2.8) |
| 28 | 15.2(2.2) | 24.1(3.3) | 3.2(0.5) | - | 40.9(3.3) |
| 2540 | 6.5(1.0) | 28.9(2.9) | 2.2(0.3) | - | 50.6(2.6) |
| 30 | 6.4(0.3) | - | 93.6(0.9) | - | - |

Table 6. Results of the correlation analysis.

| Sample | $W_{044}$ (nm) | Fe (wt.%) | $Fe^{2+}_{Oh}$ (wt.%) | $Fe^{3+}_{Th}$ (wt.%) | $Fe^{3+}_{Oh}$ (wt.%) | $Fe^{3+}_{Oh+Th}$ (wt.%) |
|---|---|---|---|---|---|---|
| 30 | 0.80 | 1.7014 | 1.5925 | 0.1089 | 0 | 0.1089 |
| 4b | 2.30 | 1.3974 | 0.7197 | 0.4625 | 0.2138 | 0.6763 |
| 4a | 3.70 | 1.0496 | 0.4114 | 0.3464 | 0.2928 | 0.6392 |
| 10a | 4.00 | 1.4838 | 0.3561 | 0.0816 | 0.3903 | 0.4719 |
| 2540 | 5.80 | 1.8362 | 0.0257 | 0.1340 | 0.3874 | 0.5214 |
| 1a | 11.09 | 1.6387 | 0.3720 | 0.4916 | 0.7751 | 1.2667 |
| 28 | 12.70 | 2.0355 | 0.0875 | 0.3969 | 0.7246 | 1.1215 |
| *R* | | 0.5551 | -0.6540 | 0.4996 | 0.9640 | 0.8931 |
| *k* | | 0.1954 | | 0.0397 | 0.0656 | 0.1053 |
| *Error* | | 0.0479 | | 0.0100 | 0.0043 | 0.0123 |

Designations:
*R* – the correlation coefficient for the linear regression $\boldsymbol{y = k*x}$, where $x = W_{044}$ (the equivalent width of the absorption band of serpentines at 0.44 μm) and *y* is equal to the quantity of Fe (in wt. %) in corresponding valency and coordination: Fe – total quantity of iron, $Fe^{2+}_{Oh}$ – ferrous octahedral iron, $Fe^{3+}_{Th}$ – ferric tetrahedral iron, $Fe^{3+}_{Oh}$ – ferric octahedral iron, $Fe^{3+}_{Oh+Th}$ – total ferric iron.

Microprobe and Mössbauer investigations of the low-iron Mg serpentine samples were undertaken, as well. Electron microprobe analysis of thin or polished sections of the samples was performed to determine their elemental composition. The measurements were made at the scanning electron microscope CamScan-4DV. The accuracy of the measurements is dependant of the content of constituents: if (C/C) > 10 wt. %, it is 2 %, if (C/C) from 5 to 10 wt. %, it is 5%, and if (C/C) from 1 to 5 wt. %, it is 10 %. However, some remarks relating the method should be made. As far as the microprobe is not sensitive to bound water and other volatiles in the layer silicates, the samples were tested for the loss of volatiles by igniting. The method of volatile content measurement in a hydrated sample is based on burning of the sample to a constant (invariable) mass. The quantity of lost volatiles corresponds to the difference in mass of the sample before and after the annealing. Necessary fractions of the samples heated up to 950 ºC have been investigated by the method. With the exception for numbers 4a and 4b which quantities were too small to get reliable data (calculated theoretically) volatile losses were determined for all samples. Then preliminary results of the microprobe analysis for the serpentine samples were recalculated and given in Table 4. Actually, microprobe measurements show that the serpentine samples are magnesian and low-iron ones.

Mössbauer spectroscopy of the samples was performed to obtain data about their $Fe^{2+}$ and $Fe^{3+}$ contents and coordinations of the ions. The measurements were carried out at a room temperature using the MS-1104E spectrometer in constant acceleration mode in absorption geometry. The $Co^{57}$(Rh) with activity of about 10 mCi was used as a γ-source. Calibration of the spectrometer was performed by the α–Fe standard technique. Processing and analysis of the experimental data were made using the MSTools program package for the model fitting of the spectra and restoration of the spectral hyperfine parameter distribution functions. The determined relative quantities of $Fe^{2+}$ and $Fe^{3+}$ (octahedral and/or tetrahedral) in reference to the values of total iron in the samples are given in Table 5. An assumption was made (Busarev et al., 2004) that the absorption band may be characteristic of the serpentine samples and deserves a special attention.

We performed also laboratory reflectance measurements of the analog samples. At standard experimental conditions (absence of an external diffused light), there is no a background in an absorption band. For this purpose, a special attention has been given to the issue every measurement. By determination, an absorption band arises in a reflectance spectrum of a sample as a difference between the measured spectral brightness and the nearest continuum. Continuum of an absorption band in a reflectance spectrum was removed by its normalization to the continuum of the spectrum after removal of the continuum slope. The relative intensity of an absorption band or its equivalent width was determined by this method. Then, the equivalent width (or an area under the drawn arbitrary continuum) of the band was calculated for the normalized reflectance spectrum of each sample according to formula (2):

$$W_{044} = \sum_{i=1}^{N} (1 - r(\lambda_i))\Delta\lambda \,, \tag{2}$$

$W_{044}$ is the equivalent width, $\Delta\lambda$ is the spectral step, $r(\lambda_i)$ are the residual intensities in the normalized spectrum, and $N$ is the number of points in the band. Thus, corrected absorption bands at 0.40-0.48 μm for 28, 1a, 2540, 10a, 4b and 4a serpentine samples are shown in Figure 6. The values of the relative intensity and other parameters of the features are listed in Table 2. We noted that values of the equivalent width of an absorption band at 0.40-0.48 μm in the reflectance spectra of considered serpentine samples form a common row. The values were disposed in the order of increase and compared with values of iron content in the corresponding samples. We determined quantities of Fe, $Fe^{2+}_{Oh}$, $Fe^{3+}_{Th}$, $Fe^{3+}_{Oh}$, and $Fe^{3+}_{Oh+Th}$ in the samples in weight percents from the microprobe and Mössbauer data (Tables 4 and 5) and included them in Table 6 for correlation analysis. Then, linear regression analysis for the dependence between the equivalent width of 0.44-μm absorption band ($W_{044}$) in the reflectance spectra and Fe (total), $Fe^{2+}_{Oh}$, $Fe^{3+}_{Th}$, $Fe^{3+}_{Oh}$, and $Fe^{3+}_{Oh+Th}$ contents in the low-iron (~1-2 wt.%) serpentine samples was performed. Results of that are given in Table 6.

Thus, we have found a middle $W_{044}$ – Fe (total) correlation ($R$ = 0.555), absence of $W_{044}$ – $Fe^{2+}_{Oh}$ correlation ($R$ = -0.654), an intermediate $W_{044}$ – $Fe^{3+}_{Th}$ correlation ($R$ = 0.500), and high correlations of $W_{044}$ – $Fe^{3+}_{Oh}$ ($R$ = 0.964) and $W_{044}$ – $Fe^{3+}_{Oh+Th}$ ($R$ = 0.893) in the low-iron Mg serpentine samples. The last two linear-proportional dependences can be expressed as the next:

$$N(Fe^{3+}_{Oh})(\text{wt. \%}) = 0.066 \times W_{044} \quad (3)$$

$$N(Fe^{3+}_{Oh+Th})(\text{wt. \%}) = 0.105 \times W_{044} \quad (4)$$

Equation (4) can be considered as a more universal one as obtained for $Fe^{3+}$ in both coordinations. We suppose that the relation may be true not only for low-iron Mg serpentines but also for other low-iron $Fe^{3+}$ compounds. It follows from theoretical and experimental investigations of $Fe^{3+}$ - content oxides and hydrated oxides (Sherman 1985; Sherman and Waite, 1985; Bakhtin, 1985). However, the assumption should be verified on a larger number of samples of low-iron $Fe^{3+}$ compounds.

At the same time, we did not find any noticeable absorption band at 0.40-0.48 μm in reflectance spectra of intermediate-iron (~5-8 wt.%) lizardite-chrysotile and antigorite samples. As it is also known from studies of serpentines and other phyllosilicates with elevated content of iron, there are only minor absorption features in their reflectance spectra in the range (Calvin and King, 1997; Hiroi and Zolensky, 1999). Based on the results, we conclude that the mechanism of light absorption due to electronic ECP-transitions in exchangeable $Fe^{3+}$ – $Fe^{3+}$ pairs is partially or completely blocked in the case of intermediate to high iron contents. For these reasons, equation (4) can not be used for a quantitative estimation of $Fe^{3+}$ content at remote investigations of solid celestial bodies.

Absorption bands of (145) Adeona, (704) Interamnia, (779) Nina, and the serpentine samples centered at 0.44-0.46 μm are compared in Fig. 6. They may be considered as the same absorption band which position is slightly changing because of structural and coordination differences in the crystal structure of the matter. Noteworthy, the band of asteroids is lower in relative intensity than that of serpentines. It may be an indication of an admixture of either different matter or similar one but with higher iron content. At the same time, such a band is absent in reflectance spectra of carbonaceous chondrites (Fig. 10). There are also two subtle absorption features in the range 0.36-0.40 μm in reflectance spectra of some serpentine samples (Fig. 11, Table 2) coinciding with that of the asteroids (Fig. 7). Therefore, we suggest that the surface matter of Adeona, Interamnia, and Nina is a mixture of carbonaceous chondrites, hydrated silicates of serpentine type and $Fe^{3+}$-content oxides and/or hydroxides.

**5. Discussion**

S-type of Pomona is confirmed by the results of our observations. We registered also weak absorption bands at 0.49-0.55 and 0.73-0.77 μm, which may be explained by the presence of the hetero-valent $Fe^{2+}$ and $Fe^{3+}$ ions in Pomona's surface matter. The spectral features could originate from electronic *d-d*-transitions and are strengthened by ECP-transitions between pairs of $Fe^{3+}$-cations and/or at $Fe^{2+} \rightarrow Fe^{3+}$ IVCT-transitions. The absorption bands slightly changing with asteroid rotation may point out to variations in the relative contents of the $Fe^{2+}$ and $Fe^{3+}$ ions and, therefore, in the oxidation state of Pomona's surface matter.

Spectral similarity of the obtained reflectance spectra of (145) Adeona, (704) Interamnia, and (779) Nina of close taxonomic types is expressed in the presence of nearly coinciding absorption features at 0.36-0.39, 0.40-0.47, and 0.50-0.85 μm. The first two of them are likely spectral features of alone ferric iron. The most prominent common spectral feature of the asteroids and their analogs, CI – CM carbonaceous chondrites and serpentines, is a wide absorption band at ~0.55-1.00 μm (Figs. 2-4, 10 and 11). It originates possibly due to $Fe^{2+} \rightarrow Fe^{3+}$ IVCT-transitions (e. g., Platonov, 1976; Bakhtin, 1985; Burns, 1993). The spectral similarity of carbonaceous chondrites and phyllosilicates (including serpentines) is explained by predominance of the latter (up to ~90 wt.%) in the matrix content of the first (e. g., Dodd, 1981). Reflectance spectra of CI – CM carbonaceous chondrites in the range 0.3-2.5 μm extensively recently studied by Cloutis et al. (2011a, 2011b). It is important to note that the measurement were performed at different phase angles. The main conclusion of the works about spectral similarity of carbonaceous chondrites and phyllosilicates is in accordance with ours.

We undertook additional investigations of low-iron Mg serpentine samples with an absorption band at 0.40-0.48 μm. Microprobe and Mössbauer measurements of the samples showed that there is a strong correlation between the equivalent width of the band and $Fe^{3+}$ content. This confirms likely the proposed mechanism of the band origin for electronic ECP-transitions in exchangeable $Fe^{3+} – Fe^{3+}$ pairs (Sherman, 1985; Sherman and Waite, 1985). We found that the highest intensity of the band is reached only in reflectance spectra of low-iron (no more then ~3 wt. % of FeO) Mg serpentines. At the same time, $Fe^{3+}$ absorption band at 0.40-0.48 μm is absent or has a low intensity in reflectance spectra of more ferrous serpentine samples. We conclude that the electronic mechanism of the light absorption is partially or completely blocked at higher iron contents. Thus, in general, the absorption feature can be used only as a qualitative indicator of ferric iron in oxidized and hydrated low-Fe compounds on the surface of asteroids.

However, an absorption band centered at 0.44-0.46 μm in reflectance spectra of the asteroids and serpentines is absent in spectra of carbonaceous chondrites (Figs. 6 and 10). The result is confirmed by measurements of other authors (e. g., Hiroi and Zolensky, 1999; Cloutis et al., 2011a, 2011b). It is probably the result of a considerably higher ferrous content in the latter. The values of FeO contents in the matrix of our three carbonaceous chondrite samples, Orguel (CI), Mighei (CM2), and Murchison (CM2) are 22.08, 27.21, and 33,43 wt. %, respectively (Zolensky et al., 1993). From a lower intensity of $Fe^{3+}$ band at 0.44-0.46 μm in reflectance spectra of studied primitive asteroids then in those of Mg serpentine samples, the supposition is made that the surface matter of the asteroids is a mixture of carbonaceous chondrites, hydrated silicates of serpentine type and $Fe^{3+}$-content oxides and/or hydroxides. This implies similar physical-chemical conditions of origin and/or evolution of the bodies.

It should be also considered a possibility of influence of G-band of F-G-stars used as solar analogs (e. g., Hardorp, 1980) on the $Fe^{3+}$ -band in reflectance spectra of asteroids. As seen in Figure 12, spectral position of the G-band (centered at 0.43 μm) overlaps partly with the $Fe^{3+}$ -band. A spectrum of our solar analog (HD10307) is shown in Figure 12, as well. Noteworthy, its spectral type is very close to that of the Sun (G2V). Taking into account the used methodology of asteroid reflectance spectrum calculation, we conclude that the influence of G-band is possible in the case of spectral differences between a solar analog and the Sun itself. To avoid this we need to choose solar analogs more carefully according to their spectral characteristics.

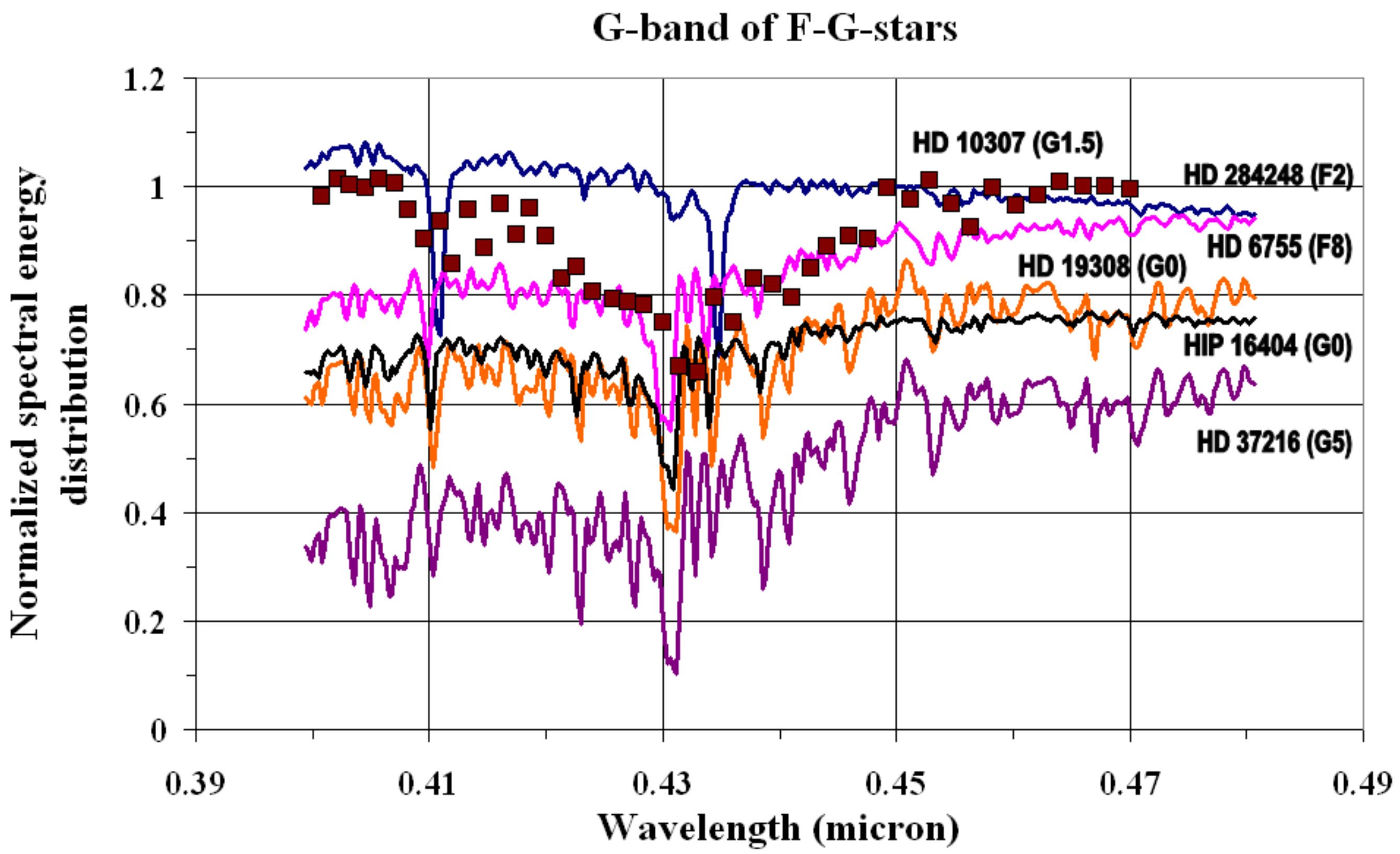

Fig. 12. Absorption G-band in spectra of some F-G– stars including HD 10307 used in the work as a solar analog.

From our spectrophotometric observations and spectral data from the SMASSII survey and asteroid taxonomy (Bus and Binzel 2002a, b), 330825 and 2012 QG42 types are assessed by us as C or S and S or B, respectively. At the same time, previous photometric and spectral observations of 330825 and 2012 QG42 showed that both bodies are of S-type (Warner et al. 2013; Taradii et al. 2013). From the published results, we conclude that our measurements could reflect some surface heterogeneity of the asteroids, especially for 2012 QG42. At the same time, reflectance spectra of the both NEAs are featureless compared to the considered main belt asteroids. Among possible reasons, we need to emphasize two common. It may be due to action of weathering factors, such as higher surface temperatures and a stronger darkening by the solar wind and UV- radiation. For instance, in the case of presence of frozen and/or adsorbed water ice and volatile organics in the asteroid surface matter, the approaching of the bodies to the Sun should lead to increasing sublimation of frozen volatiles and the formation of dark organic films on the regolith particles. At the reflection of solar light from the asteroid surface, the dark films could preclude the formation of a diffuse component containing information on the matter composition.

Discovering for the fist time apparent spectral signs of sublimation activity on (145) Adeona, (704) Interamnia, and (779) Nina deserve a special consideration. The asteroids are primitive and low-albedo, though Nina's geometric albedo is twice higher. It suggests either existence of a native water ice under the surface preserved since asteroid formation or, on the contrary, buried at previous collisional events and recently resurfaced ice (e. g., Fanale and Salvail, 1989; Jewitt and Guilbert-Lepoutre, 2012). Interamnia and Nina were in time of the observations near the perihelion distance to the Sun and their surface temperatures aroud the subsolar point reached a maximum (238.6 K and 257.3 K, respectively). Approaching to the Sun Adeona was near a middle heliocentric distance but its subsolar temperature turned out to be very close to that of Interamnia. We calculated also heliocentric distances of Adeona, Interamnia, and Nina on moments of its spectral observations by other authors. Adeona's heliocentric distance was considerably less only at observations made by McCord and Chapman (1975a) then that in ours, but we see no indications of discussed features their data (Fig. 2a). Published reflectance spectra of Interamnia (Bus and Binzel, 2002a; Lazaro et al., 2004; Fornasier et al., 2014) were obtained close to its aphelion distance to the Sun. Finally, published spectral data on Nina (Bus and Binzel, 2002a; Lazaro et al., 2004) were measured around its middle heliocentric distance to the Sun. We conclude that reflectance spectra of discussed asteroids at perihelion distances were not previously obtained or just rejected because of their dissimilarity with preceding ones. For example, Bus and Binzel (2002a) used a method of real-time examination for each registered asteroid spectrum (quick-look reduction and dividing by the spectrum of analog star) to estimate its quality.

To explain unusual reflectance spectra of some distant asteroids by the process of ice sublimation, Carvano and Lorenz-Martins (2009) for a model of a spherical asteroid reflecting light according Hapke's formula and surrounded by a faint coma of ice particles (consisting of a mixture of $H_2O$ and tholins) calculated combined reflectance spectra of such model system at different concentrations of 0.2-micron particles in the coma. We show one of their spectra corresponding to a slightly elevated dust density (2e+16 $m^{-1}$ concentration particles) in the coma (Fig. 13). One can see that shape of the spectrum in the range ~0.4-0.6 μm is very close to those of Adeona, Interamnia, and Nina (Figs. 2-4). We consider this a strong background for discovering their sublimation activity. Obviously, the asteroids should be more thoroughly investigated in the near future.

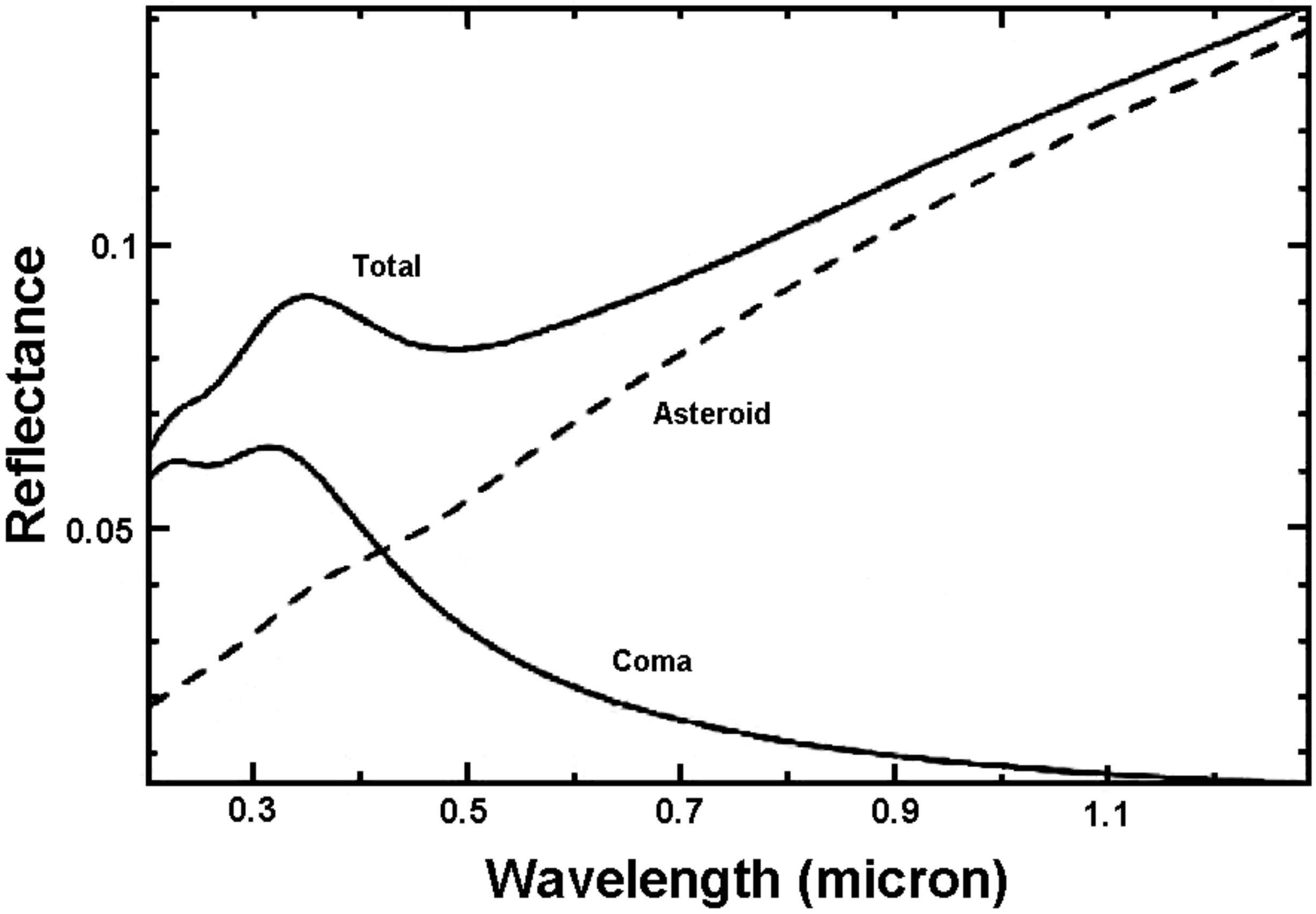

Fig. 13. Reproduced with permission after insignificant changes figure with model reflectance spectra of an asteroid covered with 10-μm particles of water ice and tholins mixture ("asteroid"), a spherical coma consisting of 0.2-μm particles of the same material (2e+16 $m^{-1}$ particle concentration) surrounding the asteroid ("coma"), and the combined system ("total") (Carvano and Lorenz-Martins, 2009).

**6. Summary and conclusions**

We obtained and analyzed reflectance spectra of different asteroids, namely: S-type (32) Pomona with predominantly high-temperature mineralogy, (145) Adeona, (704) Interamnia, and (779) Nina with predominantly low-temperature mineralogy, (330825) 2008 XE3 and 2012 QG42 which mineralogy should be specified. S-type of Pomona is confirmed by the results of our observations. At the same time, we registered weak absorption bands at 0.49-0.55 and 0.73-0.77 μm explained by the presence of the hetero-valent $Fe^{2+}$ and $Fe^{3+}$ ions in Pomona's surface matter. Found spectral similarity of the obtained reflectance spectra of (145) Adeona, (704) Interamnia and (779) Nina of close taxonomic types is expressed in the presence of nearly coinciding absorption features at 0.36-0.39, 0.40-0.47, and 0.50-0.85 μm. The absorption bands are predominantly manifestations of the electronic crystal-field *d-d*-transitions in $Fe^{3+}$ ions strengthened by transitions in magnetic exchange-coupled pairs (ECP) of neighboring cations, $Fe^{2+}$ - $Fe^{3+}$ and/or $Fe^{2+}$ - $Fe^{3+}$ (Rossman, 1975; Sherman, 1985; Sherman and Waite, 1985). The most prominent common spectral feature of the asteroids Adeona, Interamnia, and Nina and their analogs, such as CI – CM carbonaceous chondrites and serpentines, is a wide absorption band at ~0.55-1.00 μm (Figs. 2-4, 10 and 11). It is probably due to $Fe^{2+} \rightarrow Fe^{3+}$ IVCT-transitions (e. g., Platonov, 1976; Bakhtin, 1985; Burns, 1993). Thus, the carbonaceous chondrites and serpentines may be considered as the best spectral analogs for the studied primitive asteroids. A considerable fraction of the observed asteroids could be covered with these materials. This implies similar physical-chemical conditions of the origin and/or evolution of the asteroids.

We investigated more thoroughly seven samples of low-iron Mg serpentine having an absorption band at 0.40-0.48 μm which is similar to that of Adeona, Interamnia, and Nina. Microprobe and Mössbauer measurements of the samples showed that the equivalent width of the band has a strong correlation with $Fe^{3+}$ content. The relative intensity of the band is the highest in reflectance spectra of low-iron (no more than 3 wt. % of FeO) serpentines. In addition, we found experimental evidence that the absorption feature is weak or absent in reflectance spectra of serpentine samples having higher iron content. This means that the electronic mechanism of the band is partially or entirely blocked in the case. Therefore, the absorption feature can be used only as a qualitative indicator of ferric iron on the surface of asteroids and other atmosphereless celestial bodies. It is important to note that the band of considered asteroids is lower in relative intensity than that of serpentines (Fig. 6). It may point to presence in the asteroid surface matter either an admixture of spectrally different compounds or similar ones but with higher iron content.

We observed asteroids (704) Interamnia, and (779) Nina near perihelion heliocentric distances and (145) Adeona about its middle heliocentric distance and found for the first time spectral signs of a comet activity on their surface. It is manifested likely in the form of predominantly $H_2O$ ice (buried previously and resurfaced by recent impacts) sublimation and subsequent scattering by submicron ice particles of reflected sunlight in the short-wavelength range. It seems such scattering does not change position of intrinsic asteroid absorption bands but exaggerates their intensity.

Taxonomic types of 330825 and 2012 QG42 are assessed as C(S) and S(B), respectively. This is only in partial agreement with previous photometric and spectral observations of 330825 and 2012 QG42 showed that both bodies are of S-type (Warner et al. 2013; Taradii et al. 2013). To specify mineralogy of the asteroids, their additional investigation in a wider spectral range is desirable. Due to differences in geometric observational parameters, our measurements could characterize some surface heterogeneity of the asteroids, especially for 2012 QG42 as a slower rotator. However, it should be noted that reflectance spectra of the NEA asteroids are featureless compared to the main-belt asteroids (see Figs. 2-4, 8 and 9). Among possible factors, which could be responsible for the effect, we suggest two of them of most relevance and importance, namely higher surface temperatures and a more intensive darkening the surface matter by the solar wind and UV-radiation at lower heliocentric distances. In particular, dark organic films could be formed on the regolith particles on the surface of asteroids while approaching the Sun to about 1 AU or less. Indeed, the solar heating is able to substantially increase sublimation of buried or adsorbed ices and volatile organics, if any. The resulting dark organic films on silicate particles must preclude formation of diffuse component in the reflected light transmitted through the asteroid surface matter. If so, it complicates NEAs' taxonomic classification.

Probably, the most interesting results of our work are discovered spectral signs of sublimation activity of Adeona, Interamnia, and Nina at shortest heliocentric distances. Given their primitive nature, we suppose that such a cometary-like activity may be a widespread phenomenon among C and close type asteroids. Moreover, as we stated above, any type asteroids accumulated ices beneath the surface over their evolution could demonstrate such activity.

**Acknowledgments**

Microprobe measurements of the laboratory serpentine samples and microscope identification of minerals in the sample were performed by Vilen I. Feldman (died), Geological department of Lomonosov Moscow State University, Moscow, Russia.

Reflectance measurements of the carbonaceous chondrite and serpentine samples were performed by Mikhail N. Taran, Institute of Geochemistry, Mineralogy and Ore Formation, Academy of Sciences of Ukraine, Kiev, Ukraine.

Laboratory data on volatile loses in the serpentine samples were obtained by Andrey N. Grigoriev, Chemical department of Lomonosov Moscow State University, Moscow.

Mössbauer measurements of the laboratory serpentine samples were made by Mikhail V. Volovetskij, Division of Mossbauer Spectroscopy of Physical Department of Lomonosov Moscow State University, Moscow.

The authors thank the Committee of meteorites (Vernadsky Institute of Geochemistry and Analytical Chemistry, RAS, Moscow) for the provided samples of carbonaceous chondrites, as well as V. V. Nasedkin and V. A. Pavlov (Institute of Geology, Petrography, Mineralogy and Geochemistry of Ore Deposits, RAS, Moscow), D. I. Belakovskij (Fersman Mineralogical Museum, Russian Academy of Sciences, Moscow), and N. V. Skorobogatova (All-Russia Research Institute of Mineral Resources (VIMS), Moscow) for kind supply of the serpentine samples.

The observations of asteroids at the Terskol Oservatory were supported by the Russian Foundation for Basic Research (project no. 12-02-90444-Ukr).

The authors also thank the anonymous Reviewers for useful comments and helpful information.